\documentclass[reprint,aps,prb,twocolumn,superscriptaddress]{revtex4-2}

\usepackage{graphicx}
\usepackage{amsfonts}
\usepackage{amsmath}
\usepackage{amssymb}
\usepackage{bbm}
\usepackage{blindtext} 
\usepackage{placeins}
\usepackage{bm}
\usepackage{braket}
\usepackage{bbold}
\usepackage{color}
\usepackage{placeins}
\usepackage{comment}
\usepackage{hyperref}
\usepackage{latexsym}
\usepackage{multirow}
\usepackage{subfigure}
\usepackage{soul}
\usepackage{tikz}
\usepackage{tabularx}
\usepackage{upgreek}
\usepackage{verbatim}
\usepackage{xcolor}

\hypersetup{
    colorlinks=true,
    linkcolor=blue,
    urlcolor=blue,
    filecolor=blue,
    citecolor=blue
    }

\begin{document}

\title{Simulating superconductivity in mixed-dimensional $t_\parallel$-$J_\parallel$-$J_\perp$ bilayers\\ with neural quantum states}

\author{Hannah Lange}
\email{hannah.lange@lmu.de}
\affiliation{Department of Physics and Arnold Sommerfeld Center for Theoretical Physics (ASC), Ludwig-Maximilians-Universit\"at M\"unchen, Theresienstr. 37, M\"unchen D-80333, Germany}
\affiliation{Max-Planck-Institute for Quantum Optics, Hans-Kopfermann-Str.1, Garching D-85748, Germany}
\affiliation{Munich Center for Quantum Science and Technology (MCQST), Schellingstr. 4, M\"unchen D-80799, Germany}

\author{Ao~Chen}
\affiliation{Division of Chemistry and Chemical Engineering,
California Institute of Technology, Pasadena, California 91125, USA}
\affiliation{Center for Computational Quantum Physics, Flatiron Institute, New York 10010, USA}

\author{Antoine~Georges}
\affiliation{Coll\`ege de France, Universit\'e PSL, 11 place Marcelin Berthelot, 75005 Paris, France}
\affiliation{Center for Computational Quantum Physics, Flatiron Institute, New York 10010, USA}
\affiliation{Centre de Physique Théorique, Ecole Polytechnique, CNRS, Institut Polytechnique de Paris, 91128 Palaiseau Cedex, France}
\affiliation{DQMP, Université de Genève, 24 quai Ernest Ansermet, CH-1211 Genève, Suisse}

\author{Fabian~Grusdt}
\affiliation{Department of Physics and Arnold Sommerfeld Center for Theoretical Physics (ASC), Ludwig-Maximilians-Universit\"at M\"unchen, Theresienstr. 37, M\"unchen D-80333, Germany}
\affiliation{Munich Center for Quantum Science and Technology (MCQST), Schellingstr. 4, M\"unchen D-80799, Germany}

\author{Annabelle~Bohrdt}
\affiliation{Department of Physics and Arnold Sommerfeld Center for Theoretical Physics (ASC), Ludwig-Maximilians-Universit\"at M\"unchen, Theresienstr. 37, M\"unchen D-80333, Germany}
\affiliation{Munich Center for Quantum Science and Technology (MCQST), Schellingstr. 4, M\"unchen D-80799, Germany}

\author{Christopher Roth}
\affiliation{Center for Computational Quantum Physics, Flatiron Institute, New York 10010, USA}

\date{\today}
\begin{abstract}
Motivated by the recent discovery of superconductivity in the bilayer nickelate  La$_3$Ni$_2$O$_7$ (LNO) under pressure, we study a mixed-dimensional (mixD) bilayer $t_\parallel$-$J_\parallel$-$J_\perp$ model, which has been proposed as an effective low-energy description of LNO. Using neural quantum states (NQS), and in particular Gutzwiller-projected Hidden Fermion Pfaffian State, we access the ground-state properties on large lattices up to $8\times 8\times 2$ sites. We show that this model exhibits superconductivity across a wide range of dopings and couplings, and analyze the pairing behavior in detail. We identify a crossover from tightly bound, Bose-Einstein-condensed interlayer pairs at strong interlayer exchange to more spatially extended Bardeen-Cooper-Schrieffer-like pairs as the interlayer exchange is decreased. Furthermore, upon tuning the intralayer exchange, we observe a sharp transition from interlayer $s$-wave pairing to intralayer $d$-wave pairing, consistent with a first-order change in the pairing symmetry. We verify that our simulations are accurate by comparing with matrix product state simulations on coupled ladders. Our results represent the first simulation of a fermionic multi-orbital system with NQS, and provide the first evidence for superconductivity in two-dimensonal bilayers using high-precision numerics. These findings provide insight into superconductivity in bilayer nickelates and cold atom quantum simulation platforms.
\end{abstract}

\maketitle

\begin{figure}[t]
\centering
\includegraphics[width=0.49\textwidth]{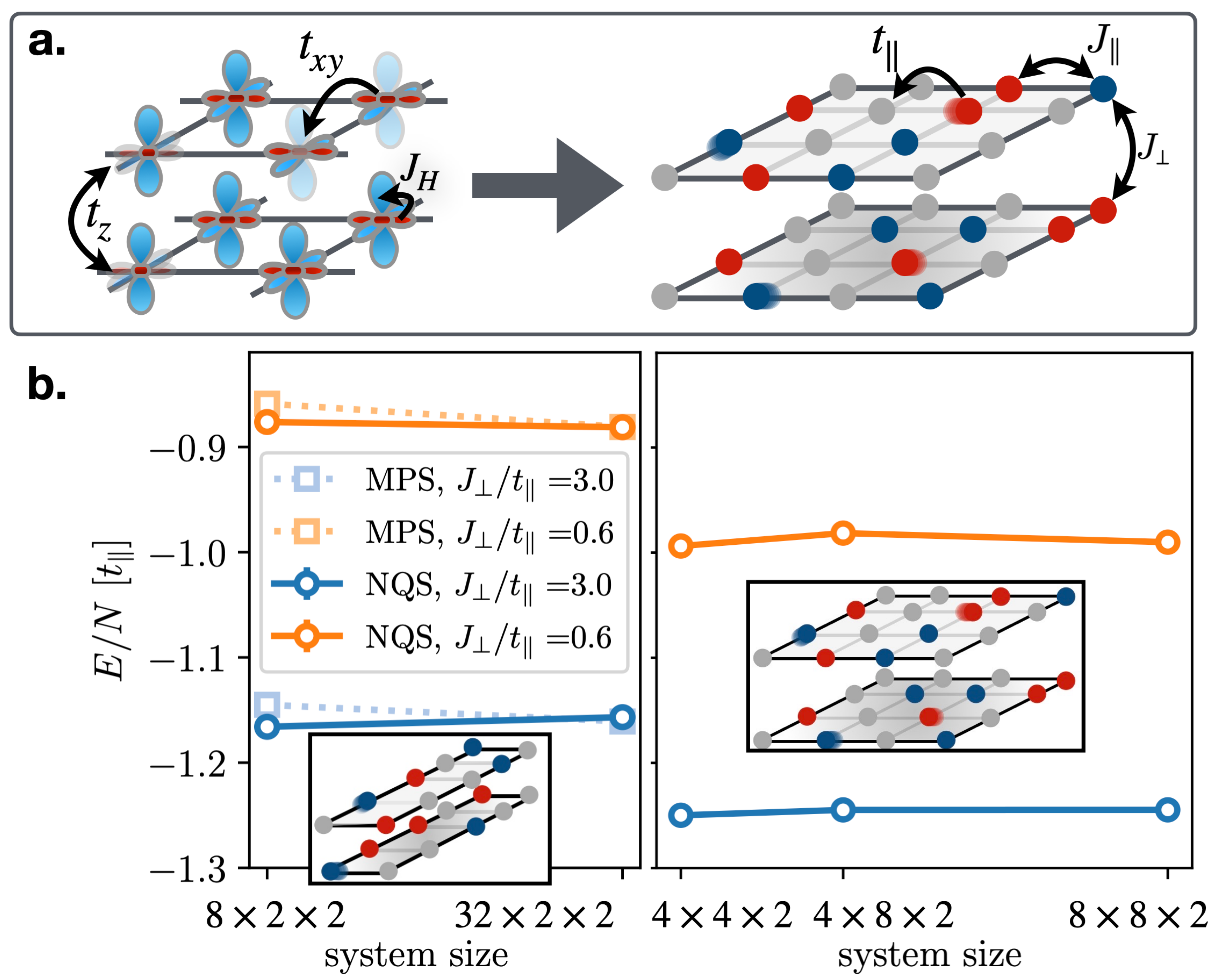}
\caption{\textbf{a.} We consider the mixed-dimensional (mixD) $t$-$J$ bilayer model (right), which can be derived from the multiband model of LNO, including $d_{z^2}$ and $d_{x^2-y^2}$ orbitals. \textbf{b.} Energies per site $E/N$ obtained with the NQS for coupled ladders, where comparisons to MPS are possible (left) and for bilayers up to size $8\times 8\times 2$. We consider two sets of parameters, with $J_\perp/t_\parallel=3.0$ and $J_\parallel/t_\parallel=0.0$ (blue) as well as $J_\perp/t_\parallel=0.6$ and $J_\parallel/t_\parallel=0.4$ (orange). For NQS (MPS), we consider periodic (open) boundaries in the long direction.
}
\label{fig:Fig1}
\end{figure}

Since the discovery of high-$T_c$ superconductivity~\cite{Bednorz1986} nearly four decades ago, extensive research efforts have revealed a rich variety of unconventional superconductors. In 2023, a superconducting transition at $T_c = 80\,\mathrm{K}$ was reported in the bilayer Ruddlesen-Popper bilayer perovskite nickelate compound La$_3$Ni$_2$O$_7$ (LNO) under applied pressure~\cite{Sun2023, Zhang2024}. Subsequent studies have shown that the physics of LNO, with $d_{z^2}$ and $d_{x^2-y^2}$ orbitals contributing at the Fermi level, is governed by strong electronic correlations \cite{Liu2024,Chgristiansson2023,Cao2024} and can be effectively captured by a mixed-dimensional (mixD) $t_\parallel$-$J_\parallel$-$J_\perp$  bilayer model with magnetic inter- and intralayer exchange $J_\perp$ and $J_\parallel$ but hopping restricted to intralayer contributions $t_\parallel$ \cite{Qu2024,Oh2023,Lu2024}, see Fig.~\ref{fig:Fig1}a. Beyond their relevance to LNO, mixD models also capture physics relevant to Kondo models and cuprate superconductors~\cite{langeKondo}.

Motivated by these developments, unbiased numerical simulations of mixed-dimensional bilayers constitute an important step toward understanding the pairing mechanism in these systems and, more broadly, the development of a theoretical description of unconventional superconductivity.
In the regime of dominant interlayer coupling $J_\perp$ and low hole-doping, pairing in mixD bilayers is well understood and arises from the interplay of the kinetic energy of doped holes and the magnetic background consisting of singlets formed on the rungs of the bilayer: In contrast to isolated holes, pairs of holes can move through the system without disrupting this singlet background, leading to the formation of tightly-bound hole pairs with large binding energies on the order of $J_\perp$ \cite{Bohrdt2022}. As a result, pair formation can be observed up to relatively high temperatures, as demonstrated in cold-atom realizations of mixD systems \cite{Hirthe2023,granet2025superconductingpairingcorrelationstrappedion}. Away from this limit, mixD bilayers have been anticipated to host a rich phase diagram, including a crossover from a Bardeen-Cooper-Schrieffer (BCS) regime to Bose-Einstein condensation (BEC) of tightly bound pairs, as discussed for single or coupled ladders in Refs.~\cite{Bohrdt2021,lange2024pairingdome,lange2024feshbachladder,schloemer2023superconductivity}, as well as multiple pairing symmetries \cite{lu2023superconductivitydopingsymmetricmass,chen2025variationmontecarlostudy,Oh2023,Yang2024,borchia2025extendedswavepairingemergent,yang2025evolutionintralayerinterlayersuperconductivity}. Given the strong pairing energy scales, this rich phase diagram may be readily observable in cold atom simulations.
  
Nevertheless, accurate numerical studies of mixD systems remain highly challenging. Apart from mean-field and dynamical mean-field theory approaches \cite{chen2025variationmontecarlostudy,Nomura2025}, investigations of the mixD model have largely relied on matrix product state (MPS) methods, which are typically restricted to ladder geometries \cite{Shen_2023,Qu2024,lange2024feshbachladder} or coupled ladders \cite{schloemer2023superconductivity,lange2024pairingdome,yang2025evolutionintralayerinterlayersuperconductivity} and open boundary conditions. To the best of our knowledge, a fully two-dimensional numerical study of the mixD $t_\parallel$-$J_\parallel$-$J_\perp$ model that captures strong correlation effects beyond mean-field-type treatments is still lacking.

Here, we address this challenge by employing a relatively new class of variational wave functions known as neural quantum states (NQS) \cite{lange2024architecturesapplicationsreviewneural,medvidović2024neuralnetwork}, and adapting them to mixD $t_\parallel$-$J_\parallel$-$J_\perp$ bilayer systems. Specifically, we combine the Hidden Fermion Pfaffian State ansatz \cite{chen2025neuralnetworkaugmentedpfaffianwavefunctions} -- which can naturally encode charge pairing if required, e.g. in the strong-$J_\perp$ regime -- with the concept of Gutzwiller projection \cite{Gutzwiller1963,lange2024tjmodelnqs}. We demonstrate that this approach enables us to investigate the rich pairing physics of the mixD model on fully two-dimensional bilayers while explicitly accounting for strong local correlations. 

The mixD $t_\parallel$-$J_\parallel$-$J_\perp$ model we study is given by
\begin{equation}
\begin{aligned}
    \mathcal{H} = &
 -t_\parallel \sum_{\langle \mathbf{i,j} \rangle, \mu, \sigma} 
 \mathcal{\hat{P}}_G\left( \hat{c}^{\dagger}_{\mathbf{i},\mu,\sigma} \hat{c}_{\mathbf{j},\mu,\sigma} + \text{h.c.} \right)\mathcal{\hat{P}}_G\\
 &+ J_\parallel \sum_{\langle \mathbf{i,j} \rangle, \mu} 
 \left( \mathbf{\hat{S}}_{\mathbf{i},\mu} \cdot \mathbf{\hat{S}}_{\mathbf{j},\mu}
 - \frac{1}{4} \hat{n}_{\mathbf{i},\mu} \hat{n}_{\mathbf{j},\mu} \right)\\
& + J_\perp \sum_{\mathbf{i}} \left(
 \mathbf{\hat{S}}_{\mathbf{i},0} \cdot \mathbf{\hat{S}}_{\mathbf{i},1} - \frac{1}{4} \hat{n}_{\mathbf{i},0} \hat{n}_{\mathbf{i},1} \right),
 \label{eq:mixDtJ}
\end{aligned}
\end{equation}
with $\hat{c}^{(\dagger)}_{\mathbf{i},\mu,\sigma}$ the annihilation (creation) operators for fermions in layer $\mu=0,1$ and with spin $\sigma=\pm 1/2$, and $\mathbf{\hat{S}}_{\mathbf{i},\mu}$ ($\hat{n}_{\mathbf{i},\mu} $) the respective spin (density) operators. DFT calculations suggest $t_\parallel \approx 0.5\, \mathrm{eV}$, $J_\parallel \approx 0.2\, \mathrm{eV}$ and $J_\perp \approx 0.3\, \mathrm{eV}$ for LNO \cite{Qu2024}.

Our Gutzwiller-projected Hidden Fermion Pfaffian States (G-HFPS) allow to simulate ground states of fully two-dimensional (2D) mixD bilayers on lattices as large as $8\times 8 \times 2$ and with periodic boundary conditions (PBC). We study the regime of $J_\perp/t_\parallel=0.6$,  $J_\parallel/t_\parallel=0.4$ and $\delta=0.5$ doping relevant to LNO, as well as perform scans through the phase diagram of the mixD bilayer model at different dopings, observing the theoretically anticipated BEC-to-BCS crossover \cite{Bohrdt2021,lange2024feshbachladder,Yang2024} as a function of $t_\parallel/J_\perp$ and a change of pairing symmetry from interlayer $s$-wave ($s_\perp$) to intralayer $d$-wave ($d_\parallel$) as a function of $J_\parallel/t_\parallel$. To our knowledge, this is the first study of the mixD $t$-$J$ bilayer model in 2D that captures strong correlation effects beyond mean-field ansätze. Furthermore, while accurate results for fermionic single-layer systems have been achieved using determinant and Pfaffian based NQS recently \cite{lange2024tjmodelnqs,gu2025solvinghubbardmodelneural,loehr2025enhancingneuralnetworkbackflow,chen2025neuralnetworkaugmentedpfaffianwavefunctions,roth2025}, for the first time a bilayer system is simulated using NQS. We observe that, due to the strong pairing inherent to this system, the underlying pairing wave function in form of the Pfaffian becomes particularly advantageous over a determinant construction. \\

\begin{figure*}[t]
\centering
\includegraphics[width=0.95\textwidth]{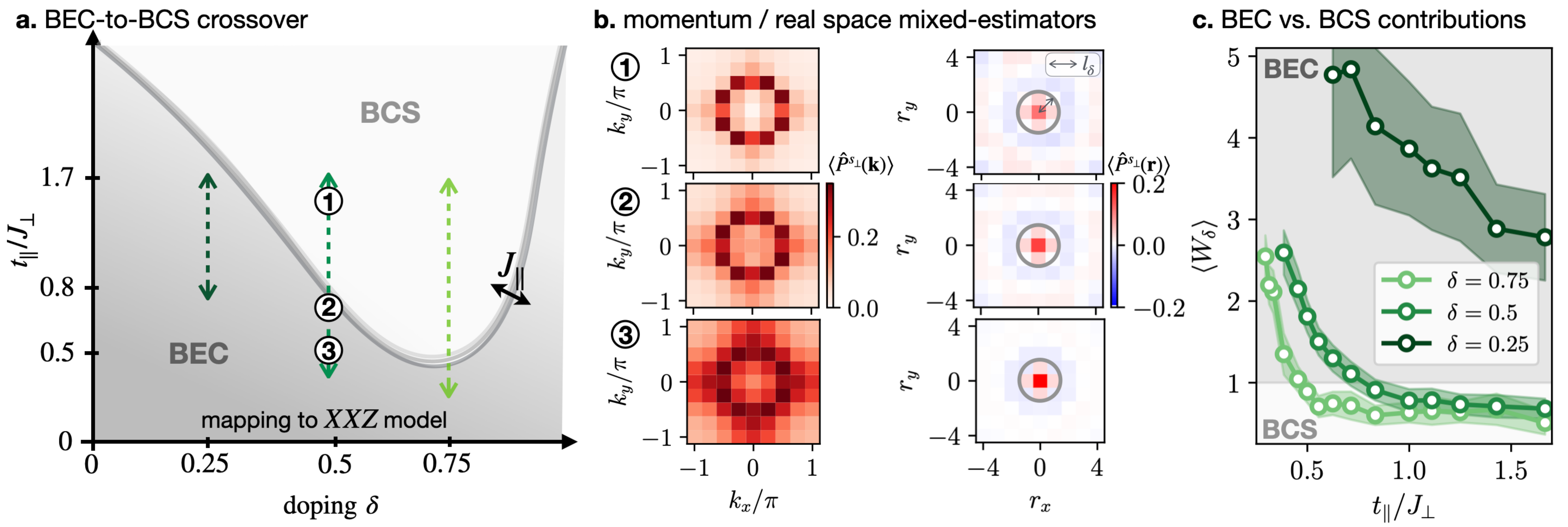}
\caption{BEC-to-BCS crossover when changing $t_\parallel/J_\perp$. \textbf{a.} Sketch of the phase diagram, with tightly bound BEC-like pairs for $t_\parallel/J_\perp \ll 1 $ and more spatially extended BCS-like pairs when $t_\parallel/J_\perp $ is increased. \textbf{b.} We show exemplary maps of the momentum-resolved pairing order parameter estimated by the mixed-estimator $\langle  \hat{P}^{s_\perp}(\mathbf{k})\rangle$ (left) and its Fourier transform $\langle  \hat{P}^{s_\perp}(\mathbf{r})\rangle$ (right) (see Eq.~\eqref{eq:MixedEstimator}). \textbf{c.} We show cuts of the BEC (BCS) condition given by $ W_\delta =\sum_{\vert \mathbf{r}\vert <l_\delta}\vert \langle  \hat{P}^{s_\perp}(\mathbf{r})\rangle \vert /\sum_{\vert \mathbf{r}\vert \geq l_\delta}\vert \langle  \hat{P}^{s_\perp}(\mathbf{r})\rangle\vert \geq 1 (<1)$ through the phase diagram at different doping levels $\delta$ as a function of $t_\parallel/J_\perp$ and at $J_\parallel/t_\parallel=0.4$. 
}
\label{fig:BECtoBCS}
\end{figure*}

\textit{Method.---} We apply a variational wave function that is conceptually similar to a mean-field pairing wave function, as used for the mixD bilayer in Ref.~\cite{chen2025variationmontecarlostudy}, but incorporate correlation effects by making use of neural networks \cite{chen2025neuralnetworkaugmentedpfaffianwavefunctions,Nomura2017,roth2025} in combination with Gutzwiller projection~\cite{Gutzwiller1963,lange2024tjmodelnqs}. To include correlations using neural networks, we expand the physical Hilbert space consisting of $N$ orbitals and $N_v$ visible fermions by additional $N_h$ hidden fermions, $N_{\mathrm{tot}}=N_v+N_h$. The pairing wave function in this expanded Hilbert space -- the Hidden Fermion Pfaffian State (HFPS) \cite{chen2025neuralnetworkaugmentedpfaffianwavefunctions} -- becomes
\begin{equation}
\psi({\bf n}) = J_{\boldsymbol{\theta}}({\bf n})\cdot  \mathrm{pf}
\begin{pmatrix} {\bf F}^{vv} & {\bf F}_ {\boldsymbol{\theta}}^{vh} ({\bf n}) \\ -{\bf F}^{vh}_{\boldsymbol{\theta}} ({\bf n})^T & {\bf F}^{hh}
\end{pmatrix},
\label{eq:HFPS}
\end{equation}
for a Fock space configuration of physical fermions, $\mathbf{n}$, and $\mathrm{pf}$ denoting the computation of a Pfaffian. The matrices ${\bf F}^{vv}$, $ {\bf F}_ {\boldsymbol{\theta}}^{vh} ({\bf n})$ and ${\bf F}^{hh}$ represent the respective couplings between visible and hidden fermions, where the configuration dependent hidden contributions are represented by a neural network with parameters $\boldsymbol{\theta}$, either by representing $ {\bf F}_ {\boldsymbol{\theta}}^{vh} ({\bf n})$ directly or by absorbing the configuration dependence of the hidden orbitals $\mathbf{F}^{hh}$ into the Jastrow factor  $J_{\boldsymbol{\theta}}({\mathbf{n}})$ without loss of generality, which tends to improve training stability \cite{chen2025neuralnetworkaugmentedpfaffianwavefunctions}. Note that $\psi$ can also express a hidden fermion determinant state (HFDS) \cite{Moreno2022} as derived in Ref. \cite{chen2025neuralnetworkaugmentedpfaffianwavefunctions}.

For the configuration dependent orbitals, we use a Group Convolutional Neural Network (GCNN), allowing us to incorporate translation, rotation, reflection and spin-parity symmetries within a single neural network forward pass \cite{SM,roth2021group}, and construct a corresponding symmetric state $\psi_\mathrm{sym}$, see SM \cite{SM}. The final wave function constructed here is given after Gutzwiller projection $\psi_\mathrm{G-HFPS}(\mathbf{n})=\hat{\mathcal{P}}_G   \psi_\mathrm{sym}(\mathbf{n})$, similar to the Gutzwiller-projected HFDS that were proposed in Ref.~\cite{lange2024tjmodelnqs}. 


Without hidden variables, Eq.~\eqref{eq:HFPS} is the mean-field solution of BCS-like Hamiltonians including pairing terms of the form $\Delta^\mathrm{MF}_{\Gamma}\hat{c}^\dagger_{\mathbf{i}}\hat{c}^\dagger_{\mathbf{j}}+\mathrm{h.c.}$ (the Pfaffian is reduced to a determinant for $\Delta^\mathrm{MF}_{\Gamma}=0$ \cite{becca_sorella_2017}). For all regimes of the phase diagram and a suitable choice of pairing symmetry $\Gamma$, we find that initialization of the HFPS such that the visible part is the solution to a Hamiltonian with a pairing field, $\Delta^\mathrm{MF}_{\Gamma}>0$, yields the best energies after the full training of the HFPS, see SM \cite{SM}. \\

\textit{Ground State Energies.---} The ground state energies obtained with the G-HFPS with $N_h=4$ and a GCNN feature dimension $f=10$ (corresponding to $\#P\approx 300,000$ trainable parameters) are shown in Fig. \ref{fig:Fig1}b for quarter-filling ($\delta=N_v/N=0.5$) and two exemplary points in the phase diagram. On the left, we compare the energies to results from MPS with total spin and charge conservation in each layer and bond dimension up to $\chi=4096$ on coupled ladders up to size $N=32\times 2\times 2$. Note that for the MPS, we are restricted to OBC in the long direction, while for the NQS we focus on PBC due to the translational symmetry inherent to the GCNN. This discrepancy leads to visibly lower energies for the smaller system but is negligibly small for the longer system: From the increase of the MPS energy from $L_x=8\to 32$, we can estimate the energy for each of the four missing bonds in $x$-direction as $c= \vert E( L_x=32)-E( L_x=8)\vert/(4\Delta L_x)\approx 0.02 \,t_\parallel$, and hence the estimated energy difference in Fig~\ref{fig:Fig1}b resulting from the different boundary choices is on the order of $4c/N=0.0007\,t_\parallel$ for the long system. Overall, the agreement with the MPS energy for these larger systems with $L_x=32$ is very good: For the system with $J_\perp/t_\parallel=3.0$ and $J_\parallel/t_\parallel=0.0$, a regime with tightly bound pairs on the rungs of the bilayer -- a relatively local state, which is favorable for the MPS representation -- the MPS energies are slightly lower with an energy difference of $\Delta E/N = 0.004\,t_\parallel$. For the regime more relevant to LNO with $J_\perp/t_\parallel=0.6$ and $J_\parallel/t_\parallel=0.4$, lower energies than the MPS energies are obtained with $\Delta E/N = -0.0002\,t_\parallel$ on the order of $4c/N$. A more detailed analysis is given in the SM \cite{SM}. The  right side of \ref{fig:Fig1}b shows that for the fully 2D systems, which are not accessible with MPS but with the G-HFPS, the energy per site $E/N$ remains approximately constant when increasing the system size. All errorbars denote errors of the mean and are typically smaller than the markers. \\

\textit{Estimation of pairing amplitudes.-- } In the following, we will investigate different pairing regimes such as a BEC-to-BCS crossover and a transition from interlayer $s$-wave ($s_\perp$) to intralayer $d$-wave ($d_\parallel$) pairing. Specifically, we consider symmetrized pair creation operators
\begin{equation}
\begin{aligned}
\hat{\Delta}_{\mathbf{i}}^{s_\perp}(\mathbf{r})&=\frac{1}{\sqrt{2}}\left(\hat{c}_{\mathbf{i},\downarrow}\hat{c}_{\mathbf{i}+(r_x,r_y,1),\uparrow} -\hat{c}_{\mathbf{i},\uparrow}\hat{c}_{\mathbf{i}+(r_x,r_y,1),\downarrow} \right)\, , \label{eq:Delta}
\\
\hat{\Delta}^{d_\parallel}_{\mathbf{i},\mu}(\mathbf{r})
&=\frac{\eta_{\mathbf{r}}}{\sqrt{2}}
\left(\hat{c}_{\mathbf{i},\downarrow}\hat{c}_{\mathbf{i}+(r_x,r_y,0),\uparrow}
-\hat{c}_{\mathbf{i},\uparrow}\hat{c}_{\mathbf{i}+(r_x,r_y,0),\downarrow}\right) \, ,
\end{aligned} 
\end{equation}
with $\ \eta_{{\mathbf{r}}\sim\hat{x}}=-\eta_{{\mathbf{r}}\sim\hat{y}}=1$ and the layer-index $\mu$ absorbed into the vectors $\mathbf{i}, \mathbf{r}$ if not stated differently. Nearest-neighbor pairs with $\mathbf{r}=(0,0,1)$ for $\Gamma=s_\perp$ and $\mathbf{r}=(\pm 1,0,0),\, (0,\pm1,0)$ for $\Gamma=d_\parallel$ will be denoted by $\hat{\Delta}^\Gamma_\mathbf{i}:= \hat{\Delta}^\Gamma_\mathbf{i}(\mathbf{r})\vert_{\vert \mathbf{r}\vert=1}$.

Superconductivity with symmetry $\Gamma=s_\perp, d_\parallel$ is characterized by particle-number symmetry breaking with order parameter $\Delta^\Gamma(\mathbf{r})=\frac{1}{N}\sum_\mathbf{i}\langle \hat{\Delta}_{\mathbf{i}}^\Gamma(\mathbf{r})+(\hat{\Delta}_{\mathbf{i}}^\Gamma(\mathbf{r}))^\dagger\rangle/2\neq 0$ or its Fourier transform $\Delta^\Gamma(\mathbf{k})=\sum_\mathbf{r}e^{i\mathbf{k}\cdot \mathbf{r}} \Delta(\mathbf{r})\neq 0$. However, in finite size systems with conserved particle number, this quantity will evaluate to zero, motivating us to study the momentum space pair correlator
\begin{align}  M^\Gamma(\mathbf{k},\mathbf{k^\prime}):=\langle\!\langle(\hat{\Delta}^\Gamma(\mathbf{k}))^\dagger \hat{\Delta}^\Gamma(\mathbf{k^\prime})\rangle\!\rangle\,,
\label{eq:M}
\end{align}
where $\langle\!\langle\dots\rangle\!\rangle$ denotes an expectation value in which contributions from unpaired particles (of the form $\langle \hat{c}^\dagger \hat{c}\rangle \langle \hat{c}^\dagger \hat{c}\rangle$) have been subtracted. Calculating the full $M^\Gamma$-matrix scales with $\mathcal{O}(N^4)$ and becomes too expensive for large system sizes. We overcome this problem by noting that for a superconductor, pair creation and annihilation processes should become fully decoupled in the thermodynamic limit and $M^\Gamma(\mathbf{k},\mathbf{k}^\prime)$ should approach rank-1 for large system sizes. In this case (as discussed in detail in Ref.~\cite{roth2025} and SM~\cite{SM}), the dominant eigenvector of $M^\Gamma(\mathbf{k},\mathbf{k}^\prime)$ corresponds to the order parameter $\Delta^\Gamma(\mathbf{k})$, that probes the linear response of the system to an applied pairing field, and can be accessed via a mixed real-momentum space correlator $\langle \hat{P}^\Gamma (\mathbf{k})\rangle$ given by
\begin{align}
     \langle  \hat{P}^\Gamma(\mathbf{k}) \rangle  
    &=\frac{1}{ \langle \!\langle (\bar{\Delta}^\Gamma)^\dagger \bar{\Delta}^\Gamma\rangle\! \rangle^{1/2}} \langle \!\langle  
    \hat{p}^\Gamma(\mathbf{k})\rangle \!\rangle, \notag\\
    \hat{p}^{\Gamma}(\mathbf{k}) &:=\hat{c}^\dagger_{{\bf k},1, \uparrow} \hat{c}^\dagger_{{\bf -k},0, \downarrow} {\hat{ \bar{\Delta}}}^{\Gamma}\label{eq:MixedEstimator},\\
    \hat{\bar{\Delta}}^\Gamma &:=\frac{1}{N}\sum_\mathbf{l}\hat{\Delta}_\mathbf{l}^\Gamma\approx \hat{\Delta}_{(L_x/2, L_y/2)}^\Gamma \notag,
\end{align}
where the last approximation renders the calculation of $\langle  \hat{P}^\Gamma(\mathbf{k}) \rangle$ tractable. We denote the corresponding Fourier transform by $\langle P^\Gamma(\mathbf{r})\rangle$. Another quantity that is often calculated to probe superconductivity in finite, particle number conserved systems are pairing correlations for real-space nearest neighbor pairs, $|{\bf r}| = 1$, by considering the long-range behavior of $\langle (\hat{\Delta}^{\Gamma}_\mathbf{i})^\dagger \hat{\Delta}^{\Gamma}_\mathbf{j}\rangle$.
In contrast to these $|{\bf r}| = 1$ pairing correlations, $\langle P^\Gamma(\mathbf{r})\rangle$ allows us to measure how tightly-bound the Cooper pairs are while, in contrast to the full computation of $M^\Gamma(\mathbf{k},\mathbf{k}^\prime)$, being computationally accessible.\\ 

To decrease the computational cost and not constrain the wave function, we use reflection, spin parity and translation symmetries from now on. To keep the number of parameters comparable to before, we set $N_h=8$ and $f=18$. The resulting relative energy error w.r.t. the fully symmetrized system is on the order of $0.1\%$, see SM \cite{SM}. Furthermore, we employ the adiabatic transport parameter update derived in Ref.~\cite{medvidovic2025adiabatictransportneuralnetwork,SM} for scans of the phase diagram at fixed $\delta$.\\

\textit{BEC-to-BCS crossover.---} 
We first study the crossover from tightly bound pairs forming a BEC at $t_\parallel < J_\perp$ to more spatially extended BCS-like Cooper pairs at large $t_\parallel>J_\perp$. More formally, we characterize the crossover by defining the average interparticle distance at doping $\delta$,
\begin{align}
    l_\delta=(1-\delta)^{-1/2}.
    \label{eq:ldelta}
\end{align}
The system is considered to be in the BEC regime as long as $l_\delta$ is larger than the size of the pairs extracted from Eq.~\eqref{eq:MixedEstimator} \cite{Bohrdt2021}.

The resulting phase diagram is sketched as a function of $t_\parallel/J_\perp$ and hole doping $\delta$ in Fig.~\ref{fig:BECtoBCS}a. In the extreme cases of very low and very high $\delta$, the crossover is shifted to large $t_\parallel/J_\perp$, which can be understood as follows: For high doping, we note that $l_\delta$ grows with the hole doping $\delta$, up to $l_{\delta\to 1}=\infty$, pushing the crossover to $t_\parallel/J_\perp\to \infty$. At low doping $\delta\approx 0$, the effective intralayer hopping $t_\parallel^\mathrm{eff}$ is suppressed as the double occupancy constraint becomes more restrictive, again increasing the value of $t_\parallel/J_\perp$ at which the crossover takes place. 

A BEC-to-BCS crossover in the mixD setting was already anticipated in \cite{schloemer2023superconductivity, Bohrdt2021}, suggested by mean-field and coupled ladder calculations, and a similar crossover was moreover studied by introducing a repulsive term on the rungs for ladder systems using MPS~\cite{lange2024feshbachladder,lange2024pairingdome,Yang2024} and for 2D systems on the mean-field level~\cite{borchia2025extendedswavepairingemergent}. However, a fully two-dimensional study including correlation effects beyond MF-like ansätze has been outstanding. 

In the following, we use the G-HFPS to study the phase diagram sketched in Fig.~\ref{fig:BECtoBCS}a. We fix $J_\parallel/t_\parallel=0.4$ and first consider $\delta=0.5$ relevant to LNO. For strong $J_\perp$, tightly bound pairs of charges form with binding energies on the order of $J_\perp$. Hence, $t_\parallel/ J_\perp\to 0$ gives rise to interlayer pairs that behave as bosonic bound states with momentum zero and hence form a BEC. In Fig.~\ref{fig:BECtoBCS}b we show the mixed-estimators in both momentum (left column) and real (right column) space. At small $t_\parallel/J_\perp\approx 0.3$ (3), we see that $\langle \hat{P}^{s_\perp}(\mathbf{k})\rangle$ is fairly dispersive, while $\langle  \hat{P}^{s_\perp}(\mathbf{r})\rangle$ decays quickly as a function of radial distance. This indicates that the pairs are de-localized in momentum space and tightly-bound in real space. By identifying rungs that are (un)occupied with a pair with an effective spin-$1/2$ up (down) particle, the system can be mapped to a XXZ model~\cite{Bohrdt2021}, with signatures for this correspondence visible already at the considered $t_\parallel/ J_\perp\approx 0.4$ in our numerics (see SM~\cite{SM}). In the opposite limit, when the intralayer hopping $t_\parallel$ dominates over $J_\perp$, the interlayer pairs become spatially extended and behave as BCS Cooper pairs \cite{Bohrdt2021}. This is visible as a peak in  $\langle \hat{P}^{s_\perp}(\mathbf{k})\rangle$ at the Fermi surface and a spatially extended, oscillating, real space pairing amplitude estimated by $\langle \hat{P}^{s_\perp}(\mathbf{r})\rangle$ in Fig.~\ref{fig:BECtoBCS}b (1) for $t_\parallel/J_\perp \approx 1.7 $. The average particle distance $l_\delta$ is indicated by the gray circles.

In the following, we will extend our analysis to different dopings $\delta=0.25, 0.5, 0.75$. For a more detailed analysis of the BEC-to-BCS crossover, we consider the real-space pairing and define the BEC and BCS regimes by the weight of $\vert \langle \hat{P}^{s_\perp}(\mathbf{r})\rangle \vert$ for $\mathbf{r}$ within and outside $l_\delta$. For the dopings considered, $l_\delta$ ranges from $l_{\delta=0.25}\approx1.15 $ to $l_{\delta=0.75}=2$. More precisely, we define the BEC (BCS) regime by $\langle W_\delta\rangle :=  \sum_{\vert \mathbf{r}\vert <l_\delta}\vert \langle  \hat{P}^{s_\perp}(\mathbf{r})\rangle \vert /\sum_{\vert \mathbf{r}\vert \geq l_\delta}\vert \langle  \hat{P}^{s_\perp}(\mathbf{r})\rangle\vert  \geq 1 $ ($\langle W_\delta\rangle<1 $). As shown in Fig.~\ref{fig:BECtoBCS}c, this condition reveals a BEC-to-BCS transition at $t_\parallel/J_\perp\approx 0.5 (0.8)$ for $\delta=0.75(0.5)$, and no transition in the considered regimes of $t_\parallel/J_\perp$ for $\delta=0.25$, in agreement with the phase diagram in Fig.~\ref{fig:BECtoBCS}a. 

Our findings demonstrate the presence of a BEC-to-BCS crossover beyond the indications for such a crossover in coupled ladders found in Ref.~\cite{schloemer2023superconductivity}, and offers insights into the microscopic origin of superconductivity in these materials: The presence of a BEC-to-BCS crossover is closely tied to a recently proposed pairing mechanism based on a Feshbach resonance between BEC and BCS-like pairing channels for both cuprate and bilayer nickelate superconductors~\cite{homeier2025feshbachcuprates,lange2024pairingdome,lange2024feshbachladder,borchia2025extendedswavepairingemergent,schloemer2023superconductivity,Yang2024}. \\

\begin{figure}[t]
\centering
\includegraphics[width=0.495\textwidth]{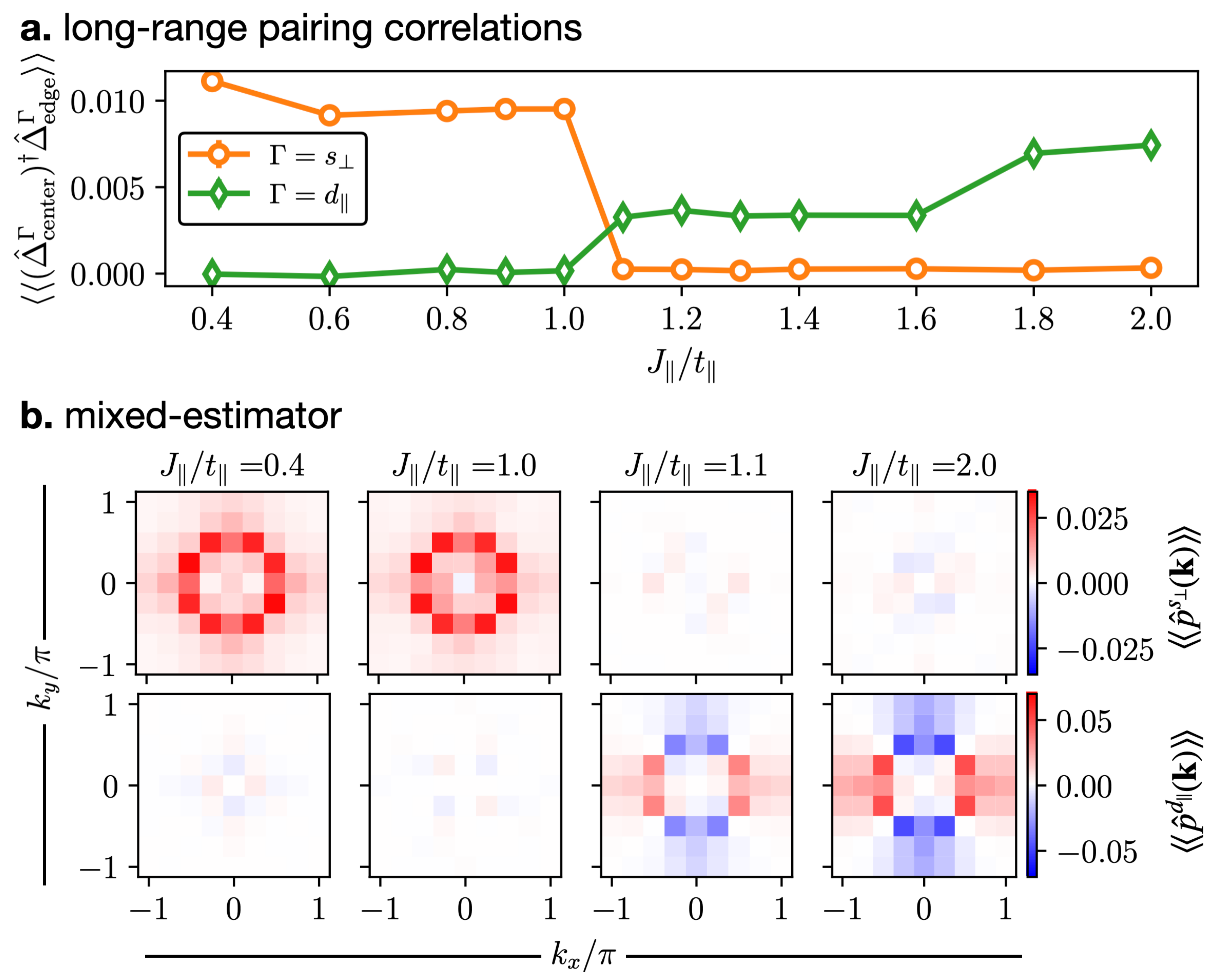}
\caption{Inter- and intralayer pairing symmetries when changing $J_\parallel/t_\parallel$ for fixed $J_\perp/t_\parallel=0.6$ and $\delta=0.5$. \textbf{a.} Long-range pairing correlations from the center site at $(4,4)$ and averaged over all four edges $\langle\! \langle (\hat{\Delta}^\Gamma_\mathrm{center})^\dagger \hat{\Delta}_\mathrm{edge}^\Gamma\rangle \!\rangle$ for $\Gamma=s_\perp$ (orange) and $\Gamma=d_\parallel$ (green). \textbf{b.} Momentum-resolved mixed-estimator $\langle \!\langle \hat{p}^\Gamma(\mathbf{k})\rangle \!\rangle $ for $\Gamma=s_\perp$ (top row) and $\Gamma=d_\parallel$ (bottom row).}
\label{fig:dswave}
\end{figure}

\textit{Pairing Symmetries.---} In Fig.~\ref{fig:dswave}a we show the long-range contributions to the pair-pair correlation function
\begin{align}
    \langle\!\langle (\hat{\Delta}^\Gamma_\mathrm{center})^\dagger \hat{\Delta}^\Gamma_\mathrm{edge}\rangle\rangle = \frac{1}{4}\sum_{\mathrm{e}\in \mathcal{E}} \langle\!\langle (\hat{\Delta}^\Gamma_{(L_x/2, L_y/2)})^\dagger \hat{\Delta}^\Gamma_\mathrm{e}\rangle\rangle
\end{align}
with $(\hat{\Delta}^\Gamma_{\mathbf{i}})^{(\dagger)}$ restricted to pairs with size $\vert \mathbf{r} \vert= 1$, and the system edges denoted by $\mathcal{E}=\{(0,0), (0,L_y-1), (L_x-1,0),(L_x-1,L_y-1) \}$. In agreement with Refs.~\cite{yang2025evolutionintralayerinterlayersuperconductivity,chen2025variationmontecarlostudy,Oh2023}, we observe a sudden change in the long-range pair-pair correlation symmetry when increasing $J_\parallel$ above $J^c_\parallel\approx t_\parallel$, where $\langle\!\langle (\hat{\Delta}^{s_\perp}_\mathrm{center})^\dagger \hat{\Delta}^{s_\perp}_\mathrm{edge}\rangle\rangle$ drops to zero, while $\langle\!\langle (\hat{\Delta}^{d_\parallel}_\mathrm{center})^\dagger \hat{\Delta}^{d_\parallel}_\mathrm{edge}\rangle\rangle$ increases from zero to a finite value. This sudden change is furthermore reflected in the respective mixed-estimators, $\langle \langle \hat{p}^{s_\perp }(\mathbf{k})\rangle \rangle$ and $\langle \langle \hat{p}^{d_\parallel}(\mathbf{k})\rangle \rangle$, in Fig.~\ref{fig:dswave}b, for which we also observe a change in the signal amplitude from zero to a finite value when tuning $J_\parallel$ above or below $J^c_\parallel\approx t_\parallel$. In contrast to the pair-pair correlations, the mixed-estimators also take into account contributions from more spatially extended pairs beyond nearest-neighbors, and are hence more general than the pair-pair correlations. All observations indicate a first-order phase transition at $J_\parallel^c$~\cite{yang2025evolutionintralayerinterlayersuperconductivity,chen2025variationmontecarlostudy,Oh2023}. To exclude the possibility of different pairing symmetries (like $s_\parallel$), we provide a more detailed analysis in the SM~\cite{SM}.\\

\textit{Conclusions.---} In this work, we have applied neural quantum states (NQS) -- specifically a new combination of Gutzwiller projection with the recently developed Hidden Fermion Pfaffian States (HFPS) -- to the mixed-dimensional bilayer $t$-$J$ model, motivated by the bilayer superconductor La$_3$Ni$_2$O$_7$ (LNO) and recent cold atom simulations. We have achieved the first fully 2D study of this system that includes strong correlation effects. Our results establish that NQS can reliably capture both ground-state energies and subtle pairing phenomena in fermionic bilayers, including a BEC-to-BCS crossover and a first-order transition between interlayer $s$-wave and intralayer $d$-wave pairing symmetries. These observations enable insights into the microscopic origin of pairing in the mixD model and, more generally, superconductors like LNO.
Beyond the specific application to the mixD bilayer system, this study highlights that NQS provide a powerful and flexible framework for studying strongly correlated fermionic systems and can access regimes beyond what has been explored with conventional methods. Looking ahead, the present work paves the way for the simulation of systems relevant to real materials and quantum simulation platforms using NQS.  \\

\textit{Acknowledgments.---} 
We wish to thank Annika Böhler, Kyle Eskridge, Yusuke Nomura, Peter Rosenberg and Shiwei Zhang for valuable discussions. This research was supported by the Deutsche
Forschungsgemeinschaft (DFG, German Research Foundation)
under Germany’s Excellence Strategy EXC-2111 Grant
No. 390814868 and the European Research Council
(ERC) under the European Union’s Horizon 2020 research
and innovation program (Grant Agreement No. 948141),
ERC Starting Grant SimUcQuam and ERC Grant QuaQuaMA (Grant No. 101217531). We also acknowledge the support provided by the Flatiron Institute. The Flatiron Institute is a division of the Simons Foundation. Numerical simulations were performed on the Paderborn Center for Parallel Computing and the Arnold Sommerfeld Cluster. CR acknowledges support provided by the Flatiron Institute, a division of the Simons Foundation. The G-HFPS simulations are performed using Quantax \cite{quantax}, which uses lrux \cite{chen2026lruxfastlowrankupdates} to accelerate large scale Monte Carlo. MPS calculations have been performed using the \textsc{SyTen} toolkit, developed and maintained by C. Hubig, F. Lachenmaier, N.-O. Linden, T. Reinhard, L. Stenzel, A. Swoboda, M. Grunder, S. Mardazad, F. Pauw and S. Paeckel. Information is available at \href{https://syten.eu/}{www.syten.eu}.





\bibliography{references.bib}
\widetext


\newpage
\pagebreak
\appendix
\widetext

\newpage

\begin{center}
\textbf{\large Supplementary Materials}
\end{center}

\setcounter{equation}{0}
\setcounter{figure}{0}
\setcounter{table}{0}
\setcounter{page}{1}
\makeatletter
\renewcommand{\theequation}{S\arabic{equation}}
\renewcommand{\thefigure}{S\arabic{figure}}

Throughout this Supplementary Material we consider the energy in units of the intralayer hopping $t_\parallel$.

\section{Connection of mixed-dimensional bilayers to LNO}
LNO has a bilayer structure with two nickel-oxide layers connected by vertical Ni-O-Ni bonds.
The onset of superconductivity under applied pressure of $14$ GPa is accompanied by a structural transition from the $Amam$ phase to the higher-symmetry $Fmmm$ phase, where the vertical bond angle changes from $168^\circ$ to $180^\circ$ \cite{Sun2023}. The increase in interlayer coupling during this transition suggests that stronger interlayer coupling enhances superconductivity.

Density functional theory (DFT) calculations have revealed that the relevant orbitals near the Fermi level in LNO are given by the Ni $3d_{x^2-y^2}$ and $3d_{z^2}$ orbitals, see Fig. \ref{fig:Fig1}a,
\begin{align}
    \hat{\mathcal{H}}_{2o} =& -t_{xy}\sum_{\langle \mathbf{i,j}\rangle,\mu,\sigma}\hat{\mathcal{P}}_G\left((\hat{c}^{(x^2-y^2)}_{\mathbf{i},\mu,\sigma})^\dagger\hat{c}^{(x^2-y^2)}_{\mathbf{j},\mu,\sigma}+\mathrm{h.c.}\right)\hat{\mathcal{P}}_G -t_{z}\sum_{ \mathbf{i}\rangle,\sigma}\hat{\mathcal{P}}_G\left((\hat{c}^{(z^2)}_{\mathbf{i},0,\sigma})^\dagger\hat{c}^{(z^2)}_{\mathbf{j},1,\sigma}+\mathrm{h.c.}\right)\hat{\mathcal{P}}_G\notag \\
    &+ J_{xy} \sum_{\langle \mathbf{i,j}\rangle,\mu}\hat{\mathbf{S}}^{(x^2-y^2)}_{\mathbf{i},\mu}\cdot \hat{\mathbf{S}}^{(x^2-y^2)}_{\mathbf{j},\mu}+ J_{z} \sum_{\mathbf{i}}\hat{\mathbf{S}}^{(z^2)}_{\mathbf{i},0}\cdot \hat{\mathbf{S}}^{(z^2)}_{\mathbf{j},1} -J_H \sum_{\mathbf{i},\alpha}\hat{\mathbf{S}}^{(z^2)}_{\mathbf{i},\mu}\cdot \hat{\mathbf{S}}^{(x^2-y^2)}_{\mathbf{j},\mu}, \label{eq:2orbitalmodel}
\end{align}
with the annihilation (creation) operators for the two orbitals, $(\hat{c}^{(x^2-y^2)}_{\mathbf{i},\mu,\sigma})^{(\dagger)}$ and $(\hat{c}^{(z^2)}_{\mathbf{i},\mu,\sigma})^{(\dagger)}$, and spin operators $\hat{\mathbf{S}}^{(x^2-y^2)}_{\mathbf{i},\mu}$ and $\hat{\mathbf{S}}^{(z^2)}_{\mathbf{i},\mu}$. Inter-(intra-) layer hopping of the $d_{x^2-y^2}$ ($d_{z^2}$) orbitals vanishes due to their orbital character displayed in Fig.~\ref{fig:Fig1}a. The $d_{x^2-y^2}$ orbitals are approximately quarter-filled; the $d_{z^2}$ orbitals half-filled \cite{Sun2023,Luo2023,Zhang2023Electronic,Saikabara2024}. Hence, the contribution of the $t_z$ term vanishes. The $3d$ character of the electronic structure together with the absence of perfect nesting in the non- interacting model as well as the minor role of electron-phonon coupling suggested by experiments \cite{LI2025180} and theory \cite{You2025} indicates that strong electronic correlations must be taken into account for an accurate description of LNO \cite{Liu2024,Chgristiansson2023,Cao2024}. \\

The strong interlayer interactions dominated by the almost localized $d_{z^2}$ have been shown to be promoted to the $3d_{x^2-y^2}$ orbitals through a Hund's coupling which favors a spin-triplet between the two orbitals \cite{Cao2024,Lu2024}. Integrating out the $d_{z^2}$ degrees of freedom yields a mixed-dimensional (mixD) bilayer $t_\parallel$-$J_\parallel$-$J_\perp$ model shown in Fig. \ref{fig:Fig1}a (right) and given by Eq.~\eqref{eq:mixDtJ} \cite{Qu2024,Oh2023,Lu2024}. This can be seen as follows:

In the limit of strong $J_H$, triplets form between the orbitals, and the low-energy subspace consists of the triplet states
\begin{equation}
    \begin{aligned}
    \ket{+1}_{\mathbf{i},\mu}&=\vert \underbrace{\uparrow }_{d_{z^2}\,\mathrm{orbital}}\underbrace{\uparrow }_{d_{x^2-y^2}\,\mathrm{orbital}}\rangle_{\mathbf{i},\mu}\\
    \ket{0}_{\mathbf{i},\mu}&=\frac{1}{\sqrt{2}}\left(\vert \uparrow \downarrow\rangle_{\mathbf{i},\mu}+\vert \downarrow \uparrow\rangle_{\mathbf{i},\mu}\right)
    \\
    \ket{-1}_{\mathbf{i},\mu}&=\vert \downarrow \downarrow\rangle_{\mathbf{i},\mu}.
    \end{aligned}
\end{equation}
For each rung, this results in nine low-energy states. It can be shown that the action of the operator $\hat{\mathcal{H}}_{z^2}=\hat{\mathbf{S}}^{(z^2)}_{\mathbf{i},0}\cdot \hat{\mathbf{S}}^{(z^2)}_{\mathbf{j},1}$ corresponds to the action of $\hat{\mathcal{H}}_{x^2-y^2}=\hat{\mathbf{S}}^{(x^2-y^2)}_{\mathbf{i},0}\cdot \hat{\mathbf{S}}^{(x^2-y^2)}_{\mathbf{j},1}$ for these states \cite{Lu2024}. 
This allows to integrate out the $d_{z^2}$ orbitals by replacing the $\hat{\mathcal{H}}_{z^2}$ term in Eq.~\eqref{eq:2orbitalmodel} with $\hat{\mathcal{H}}_{x^2-y^2}$.

\section{Wave function amplitudes of Hidden Fermion Pfaffian States}
In their unprojected form, the HFPS are defined as
\begin{equation}\begin{aligned}|\psi\rangle  = \frac{1}{(N_\mathrm{tot}/2)!} \Big(\sum_{p < q} \mathbf{F}^{vv}_{pq} \hat{c}^\dagger_p \hat{c}^\dagger_q+ \sum_{\tilde{p} < \tilde{q}} \mathbf{F}^{hh}_{\tilde{p}\tilde{q}} \hat{d}^\dagger_{\tilde{p}} \hat{d}^\dagger_{\tilde{q}}+ \sum_{p, \tilde{p}} \mathbf{F}^{vh}_{p\tilde{p}} \hat{c}^\dagger_p \hat{d}^\dagger_{\tilde{p}}\Big)^{{N_{\mathrm{tot}}}/{2}}|0\rangle ,\label{eq:HiddenThouless}\end{aligned}  \end{equation}
where $\hat{c}^\dagger_p$ and $\hat{d}^\dagger_{\tilde{p}}$ denote visible and hidden fermion creation operators; and $\mathbf{F}^{vv}$, $\mathbf{F}^{hh}$, and $\mathbf{F}^{vh}$ are matrices representing the respective couplings between visible and hidden fermions. The projection of Eq.~\eqref{eq:HiddenThouless} onto Fock state configurations $\ket{\mathbf{n}}=\ket{n_1,n_2,\dots ,n_N}$, $\ket{\psi}$ is done by making the hidden configurations dependent on the physical $\mathbf{n}$, i.e. $\tilde{\mathbf{n}}=\tilde{\mathbf{n}}(\mathbf{n})$. This is done by defining the operation $\star$ representing a slicing operation selecting the rows of $\mathbf{F}$ by $\mathbf{n}\star \mathbf{F}$ and columns by $ \mathbf{F}\star\mathbf{n}$ according to the occupied orbitals of $\mathbf{n}$ and defining new hidden matrices $\mathbf{F}^{vh}(\mathbf{{n}}):=\mathbf{F}^{vh}\star \mathbf{{\tilde{n}}}(\mathbf{{n}})$ and $\mathbf{F}^{hh}(\mathbf{{n}}):=\mathbf{{\tilde{n}}}(\mathbf{{n}}) \star \mathbf{F}^{vh}\star \mathbf{{\tilde{n}}}(\mathbf{{n}})$, we have
\begin{equation}
\psi(\mathbf{n}) = \langle \mathbf{n}\vert \psi \rangle = 
\mathrm{pf} \!\left(
\begin{array}{cc}
\mathbf{n} \star \mathbf{F}^{vv} \star \mathbf{n} & \mathbf{n} \star \mathbf{F}^{vh}_{\mathbf{\theta}}(\mathbf{n}) \\
-\mathbf{F}^{vh}_{\mathbf{\theta}}(\mathbf{n})^{T} \star \mathbf{n} & \mathbf{F}^{hh}_{\mathbf{\theta}}(\mathbf{n})
\end{array}
\right).
\label{eq:HFPS_SM}
\end{equation}
Hereby, $\mathrm{pf}$ represents the calculation of the Pfaffian. The hidden contributions can be rewritten are parameterized by a neural network \cite{chen2025neuralnetworkaugmentedpfaffianwavefunctions} with parameters $\mathbf{\theta}$. In this work, we use the slightly modified version, Eq.~\eqref{eq:HFPS}, where the configuration dependence of $\mathbf{F}^{hh}_{\mathbf{\theta}}(\mathbf{n})$ is absorbed into a Jastrow factor $J_{\mathbf{\theta}}(\mathbf{n})$.

\section{Symmetries in Hidden Fermion Pfaffian States}

For all calculations, we make use of the lattice symmetries of the bilayer setup.  Here, we explain how the symmetrization is done, how the symmetry sectors are chosen and how the symmetrization is made computationally efficient. We closely follow Ref. \cite{chen2025neuralnetworkaugmentedpfaffianwavefunctions} in this section.\\

For a fermionic Hamiltonian and spinless fermions, we consider the group of symmetry operations $G$, with elements $g\in G$ and $g$ representing translations, rotations, reflections etc. of the system. The respective symmetry operators $\hat{g}$ act on the Fock state basis as
\begin{equation}
    \hat{g} |\mathbf{n}\rangle = \prod_{i}^{N} \hat{c}^{\dagger}_{g\mathbf{x}_i} |0\rangle = \Pi(\mathbf{n}, g)\, |g\mathbf{n}\rangle,
\end{equation}
with $g\mathbf{n}$ the new occupation numbers after application of $g$, and $\Pi_g(\mathbf{n})= \pm 1$ is the additional sign associated with fermion permutations accompanying $g$.  

The symmetrized wave function is given by
\begin{align}
    \psi_{\mathrm{sym}}(\mathbf{n})
    &= \sum_{g} \Pi_g(\mathbf{n})\, \chi_{g}\, \psi(g\mathbf{n})
    \label{eq:symPsi}
\end{align}
with the character of the symmetry operation $\chi_{g}$. For the HFPS \eqref{eq:HFPS}, $\psi(g\mathbf{n})$ is given by
\begin{equation}
\psi(g\mathbf{n}) = 
\mathrm{pf} \!\left(
\begin{array}{cc}
(g\mathbf{n}) \star \mathbf{F}^{vv} \star (g\mathbf{n}) & (g\mathbf{n}) \star \mathbf{F}^{vh}(g\mathbf{n}) \\
-\mathbf{F}^{vh}(g\mathbf{n})^{T} \star (g\mathbf{n}) & \mathbf{F}^{hh}(g\mathbf{n})
\end{array}
\right).
\label{eq:symHFPS}
\end{equation}
As explained in the following, we use Group Convolutional Neural Networks (GCNNs) to calculate $ \mathbf{F}^{vh}(\mathbf{n})$ and $\mathbf{F}^{vh}(\mathbf{n})$, which have the advantage that $ \mathbf{F}^{vh}(g\mathbf{n})$ and $\mathbf{F}^{vh}(g\mathbf{n})$ can be obtained from a single forward pass. The Pfaffian however has to be computed for every $g$ in Eq. \eqref{eq:symHFPS} separately. In Sec. \ref{sec:Sl}, we explain how the cost can be reduced by applying translational symmetries only on the level of a sublattice unit cell.

\subsection{Group Convolutional Neural Networks (GCNNs)}

 Group Convolutional Neural Networks (GCNNs) are a generalization of CNNs that work for non-abelian symmetry groups. These are useful for lattices, which have symmetry groups with rotations, reflections, and translations, some of which do not commute. These are constructed using equivariant convolutions, i.e. linear operations that preserve the structure of a discrete symmetry group \cite{roth2021group}. This is in contrast to symmetry-averaging procedures which have been applied in various works \cite{lange2024architecturesapplicationsreviewneural}.

In this paper, we consider the symmetry group of the square lattice $G$, including translations and the $D_4$ symmetry group. A set of objects $S$ is called the $G$-space if the objects are related by symmetry transformations of $G$. Each group convolutional layer defines a linear transformation between these $G$-spaces. Given a set of feature vectors, $\{\mathbf{f}\}_S=\{\mathbf{f}(s)\vert s\in S\}$, a function is equivariant if it obeys the constraint
\begin{align}
    F(g \{\mathbf{f}\}_S) = g F(\{\mathbf{f}\}_S),
\end{align}
where $g \in G$ are the elements of the symmetry group. In short, the layer transformations must commute with the symmetry generators of the group \footnote{The actual equivariant constraint is looser, as the operation only has to preserve the structure of the group, as the set can be arbitrarily permuted. For ease of explanation we consider the case where the ordering of the set stays the same}.

Convolutional neural networks (CNNs) consist of convolutional layers with feature maps $\mathbf{f}$ and kernels $\mathbf{k}$. While the kernel $\mathbf{k}$ is slid over the input features, resulting in different translations $(x-x',y-y')$ it is multiplied with the input feature values $\mathbf{f}(x',y')$, resulting in 
\begin{align}
    C(x,y) = [\mathbf{f}*\mathbf{k}](x,y) = \sum_{x',y'} \sum_i k_i(x-x',y-y') \, f_i(x',y').
\end{align}
The index $i$ denotes the $i$-th channel of the input feature. The GCNN generalizes the CNN to act over a more general discrete group $G$, 
\begin{align}
    C(g) = \sum_{h\in G}\sum_i k_i(g^{-1}h) \, f_i(h).
\end{align}
This general group convolution is equivariant under a group operation $u$ since \cite{roth2021group} 
\begin{align}
    C(ug) = \sum_{h\in G}\sum_i k_i(g^{-1}u^{-1}h) \, f_i(h)= \sum_{h^\prime \in G}\sum_i k_i(g^{-1}h^\prime) \, f_i(uh^\prime)=uC(g).
\end{align}

\subsection{Choice of Symmetry Sectors}
\begin{figure}[htp]
\centering
\includegraphics[width=0.9\textwidth]{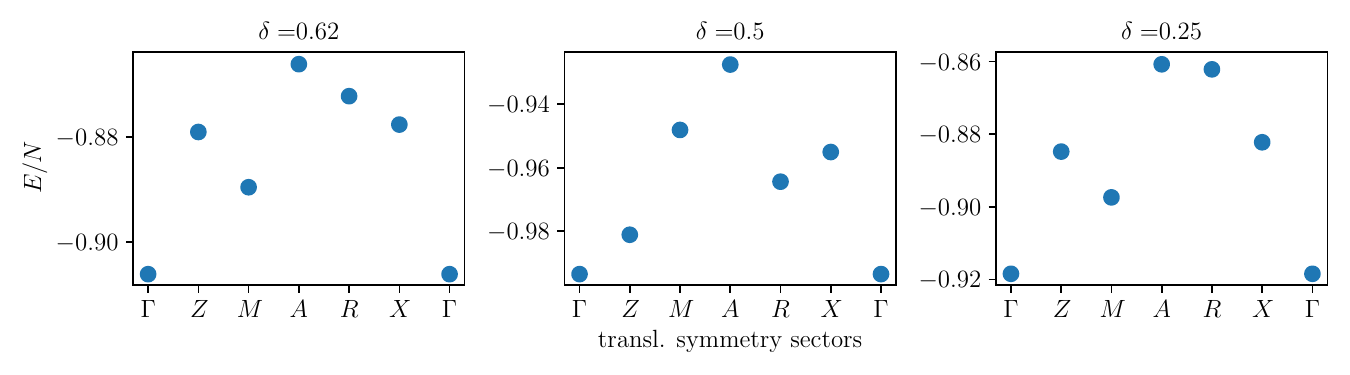}
\caption{Ground state energies per site $E/N$ for $4\times 4\times 2$ mixD bilayers obtained with the HFPS for different dopings $\delta=0.62,0.5,0.25$ (left to right) and different translational symmetry sectors indicated by the respective $k$-point in the reciprocal lattice.
}
\label{fig:symsectors}
\end{figure}

Since we make use of the translational, rotational and spin-flip symmetry of the system, we perform scans through these symmetry sectors in order to obtain the symmetry sector of the ground state. Exemplary scans of the translational symmetry sector for $4\times 4\times 2$ mixD bilayers at different dopings $\delta=0.62,0.5,0.25$ are shown in Fig. \ref{fig:symsectors}.

\subsection{Sublattice symmetries \label{sec:Sl}}

In general, Eq. \ref{eq:symHFPS} requires applying summing over $|G|$ Pfaffians. However, the number of unique Pfaffians in the sum can be reduced by imposing symmetries on the couplings between visible fermions, $F^{vv}$. For example if the couplings between electrons are translationally invariant, only point group symmetries need to be restored. In practice, we find that it is best to choose the couplings of $F_{vv}$ to have "sublattice symmetry". Here a unit cell with several different sub-lattices is defined, and translational symmetry is enforced for the sublattices but not within the unit cell. This is explained in detail in Ref. \cite{chen2025neuralnetworkaugmentedpfaffianwavefunctions}. 

\begin{figure}[t]
\centering
\includegraphics[width=0.9\textwidth]{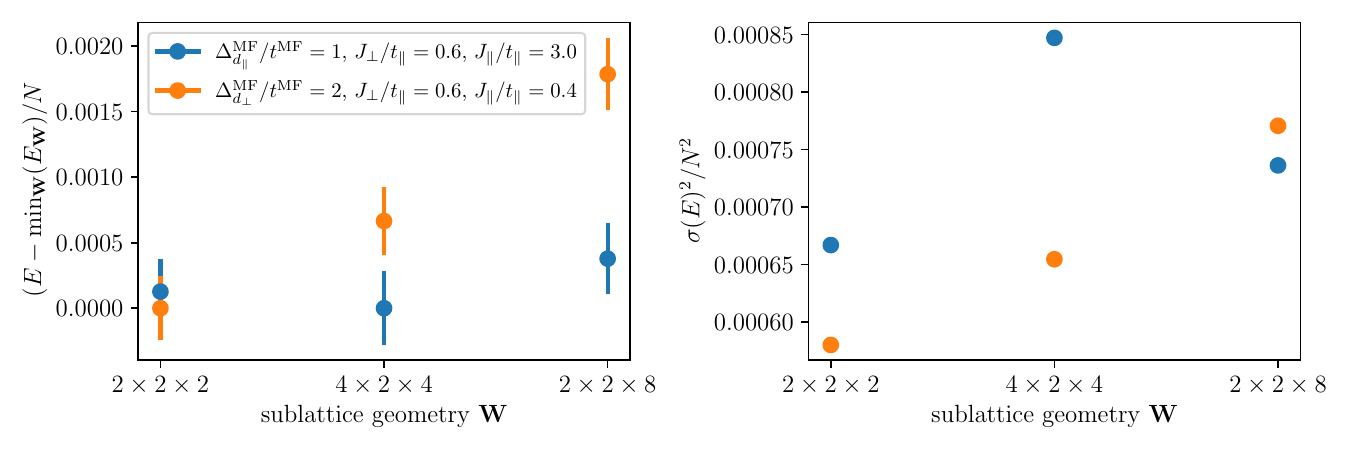}
\caption{Final ground state energies (left) and variance (right) per site $N$ for $8\times 8\times 2$ mixD bilayers obtained with the HFPS for different sublattice geometries. 
We consider hole doping $\delta=0.5$ with $J_\parallel/J_\perp=0.4, 2.0$ and different mean field pairing fields $\Delta^\mathrm{MF}_\Gamma$ ($\Gamma=s_\perp, d_\parallel)$ respectively, corresponding to states with $s_\perp$- and $d_\parallel$-wave pairing.}
\label{fig:sl}
\end{figure}

In practice, we determine the sublattice size $W_x\times W_y\times W_z$ by a scan, choosing the sublattice geometry that produces the lowest ground state energy. Exemplary scans for $8\times 8\times 2$ systems at different dopings $\delta=0.5,0.25$ and $J_\parallel/J_\perp=0.4, 3.0$ respectively -- corresponding to states with $s$- and $d$- wave pairing --, are shown in Fig. \ref{fig:sl}.

\section{Convergence of the G-HFPS}

\subsection{Optimization}
\subsubsection{Stochastic reconfiguration}
We optimize the G-HFPS using the minimum-step stochastic reconfiguration algorithm from Refs.~\cite{chen2023efficient,rende_simple_2023}, with a pseudo-inverse with relative singular value cutoff $\delta=10^{-12}$, learning rate $\eta=0.01$ and $10000$ samples in each iteration.

For the transfer learning, we use a similar algorithm that explicitly follows the respective state under consideration as a function of the Hamiltonian parameters $\mathbf{\lambda}$ fro Ref.~\cite{medvidovic2025adiabatictransportneuralnetwork}, which is summarized in the following section.

\subsubsection{Adiabatic transport}

For the scans of the phase diagram in the main text, Figs.~\ref{fig:BECtoBCS} and ~\ref{fig:dswave}, we use the adiabatic transport protocol developed in Ref.~\cite{medvidovic2025adiabatictransportneuralnetwork}, which is specifically taylored to follow an eigenstate when evolving the Hamiltonian parameters.

For a given target eigenenergy of an Hamiltonian $\mathcal{H}_{\mathbf{\lambda}}$, $\omega_{\mathbf{\lambda}}$, we estimate the target eigenstate
\(|\Psi_{\mathbf{\lambda}}\rangle\) by the sequence (starting from a state $|\Psi_0\rangle$)
\begin{equation}
    |\Psi\rangle =\lim_{k\to \infty}|\Psi_k\rangle=\lim_{k\to \infty} \left[(\mathcal{H}_{\mathbf{\lambda}} - \omega_{\mathbf{\lambda}})^{-1} \right]^k|\Psi_0\rangle=\lim_{k\to \infty} (\mathcal{H}_{\mathbf{\lambda}} - \omega_{\mathbf{\lambda}})^{-k} |\Psi_0\rangle=\lim_{k\to \infty} \sum_n c_n(E_{\mathbf{\lambda}}(\mathbf{\sigma}_n) - \omega_{\mathbf{\lambda}})^{-k} |\mathbf{\sigma}_n\rangle,
    \label{eq:ipi}
\end{equation}
converging to the ground state $|\Psi\rangle$ after a sufficient number of iterations $k$. 

For a variational state parameterized by network parameters \(\mathbf{\theta}\),
the new state after an infinitesimal change in parameters can be expanded as
\begin{equation}
    |\Psi_{\theta + \delta\theta}\rangle \approx 
    |\Psi_{\mathbf{\theta}}\rangle + \sum_{\mu} \delta\theta_\mu |\partial_\mu \Psi_{\mathbf{\theta}}\rangle,
    \label{eq:variation}
\end{equation}
where \(|\partial_\mu \Psi_{\mathbf{\theta}}\rangle = \frac{\partial}{\partial \theta_\mu}|\Psi_{\mathbf{\theta}}\rangle\). Projecting one step of Eq.~\eqref{eq:ipi} onto the tangent space of the variational manifold, multiplying from the left by \((\mathcal{H}_{\mathbf{\lambda}} - \omega_{\mathbf{\lambda}})\) and taking inner products
with the tangent vectors \(\langle \partial_\mu \Psi_{\mathbf{\theta}}|\), yields
\begin{equation}
    \sum_\nu \mathrm{Re}\!\left[\langle \partial_\mu \Psi_{\mathbf{\theta}} | (\mathcal{H}_{\mathbf{\lambda}} - \omega_{\mathbf{\lambda}}) | \partial_\nu \Psi_{\mathbf{\theta}} \rangle \right] 
    \delta\theta_\nu
    = -\,\mathrm{Re}\!\left[\langle \partial_\mu \Psi_{\mathbf{\theta}} | \mathcal{H}_{\mathbf{\lambda}} | \Psi_{\mathbf{\theta}} \rangle \right].
    \label{eq:linear_system}
\end{equation}
Equation~\eqref{eq:linear_system} defines a linear system for the parameter updates,
which we write compactly as \cite{medvidovic2025adiabatictransportneuralnetwork}
\begin{equation}
    \mathbf{G} \, \mathbf{\delta\theta} = -\mathbf{f},
    \label{eq:update}
\end{equation}
where the generalized covariance matrix \(\mathbf{G}\) and force vector \(\mathbf{f}\) are 
\begin{align}
    G_{\mu\nu} &= 2\,\mathrm{Re}\!\left[\langle \partial_\mu \Psi_{\mathbf{\theta}} | (\mathcal{H}_{\mathbf{\lambda}} - \omega_{\mathbf{\lambda}}) | \partial_\nu \Psi_{\mathbf{\theta}} \rangle \right], \\
    f_\mu &= 2\,\mathrm{Re}\!\left[\langle \partial_\mu \Psi_{\mathbf{\theta}} | \mathcal{H}_{\mathbf{\lambda}} | \Psi_{\mathbf{\theta}} \rangle \right].
\end{align}
The update is implemented by defining 
the local energy $E_{\mathrm{loc},\mathbf{\lambda}}(\mathbf{\sigma})= \frac{\langle \mathbf{\sigma} | H | \Psi_\theta\rangle}{\langle \mathbf{\sigma} | \Psi_\theta\rangle}$, the Jacobian 
\begin{align}
    J_\mu(\mathbf{\sigma}) 
    &= \frac{\langle \mathbf{\sigma} | \partial_\mu \Psi_\theta\rangle}{\langle \mathbf{\sigma} | \Psi_\theta\rangle}, 
\end{align}
and the projected Jacobian
\begin{align}
    P_\mu( \mathbf{\sigma})
    &= \frac{\langle \mathbf{\sigma} | (\mathcal{H}_{\mathbf{\lambda}} - \omega_{\mathbf{\lambda}}) | \partial_\mu \Psi_\theta\rangle}
    {\langle \mathbf{\sigma} | \Psi_\theta\rangle}=\partial_\mu E_{\mathrm{loc},\mathbf{\lambda}}(\mathbf{\sigma}) + J_\mu(\mathbf{\sigma}) \bigl(E_{\mathrm{loc},\mathbf{\lambda}}(\mathbf{\sigma}) - \omega_{\mathbf{\lambda}} \bigr)
\end{align}
Replacing $J_{i\mu} = J_\mu(\mathbf{\sigma}_i)/\sqrt{N_s}$, $P_{i\mu} = P_\mu(\mathbf{\sigma}_i)/\sqrt{N_s}$, and $\varepsilon_i = E_{\mathrm{loc}}(\mathbf{\sigma}_i)/\sqrt{N_s}$, the update equation reduces to the linear system
\begin{equation}
    J^{\top} P \, \delta\theta = -\, J^{\top} \varepsilon,
\end{equation}
which is solved similarly to the stochastic reconfiguration update \cite{medvidovic2025adiabatictransportneuralnetwork}.

\subsection{Mean-Field initialization}
We start the optimization of the G-HFPS from a (Gutzwiller projected) mean-field solution of 
\begin{align}
H &= -t^{\mathrm{MF}}\sum_{\langle i,j\rangle,\sigma} \left( c^\dagger_{i\sigma} c_{j\sigma}+\mathrm{h.c.}\right)
 + U^{\mathrm{MF}}\sum_{i} n_{i\uparrow} n_{i\downarrow}  + \sum_{i,j} \bigl(\Delta_{ij}^{\mathrm{MF}}\, c^\dagger_{i\uparrow} c^\dagger_{j\downarrow} + \text{h.c.}\bigr) .
\end{align}
We consider the following pairing fields
\begin{equation} 
\begin{aligned}
    \hat{\Delta}^{\mathrm{MF}}_{s_\perp}&\propto \left(\hat{c}_{\mathbf{l},0,\downarrow}\hat{c}_{\mathbf{l},1,\uparrow} -\hat{c}_{\mathbf{l},0,\uparrow}\hat{c}_{\mathbf{l},1,\downarrow} \right)\\
\hat{\Delta}_{d_\parallel, \mu}^{\mathrm{MF}}
&\propto\sum_{\delta=\pm\hat{x},\pm\hat{y}}{\eta_\delta^{d}}
\left(\hat{c}_{\mathbf{l},\mu,\downarrow}\hat{c}_{\mathbf{l}+\delta,\mu,\uparrow}
-\hat{c}_{\mathbf{l},\mu,\uparrow}\hat{c}_{\mathbf{l}+\delta,\mu,\downarrow}\right)
\\
\hat{\Delta}_{s_\parallel, \mu}^{\mathrm{MF}}
&\propto\sum_{\delta=\pm\hat{x},\pm\hat{y}}
\left(\hat{c}_{\mathbf{l},\mu,\downarrow}\hat{c}_{\mathbf{l}+\delta,\mu,\uparrow}
-\hat{c}_{\mathbf{l},\mu,\uparrow}\hat{c}_{\mathbf{l}+\delta,\mu,\downarrow}\right)\\
\hat{\Delta}_{s_\perp+d_\parallel}^{\mathrm{MF}} &\propto \hat{\Delta}_{s_\perp}+\hat{\Delta}_{d_\parallel, 0}+\hat{\Delta}_{d_\parallel, 1}.
\label{eq:Deltas_SM}
\end{aligned}
\end{equation}

\begin{figure}[t]
\centering
\includegraphics[width=0.7\textwidth]{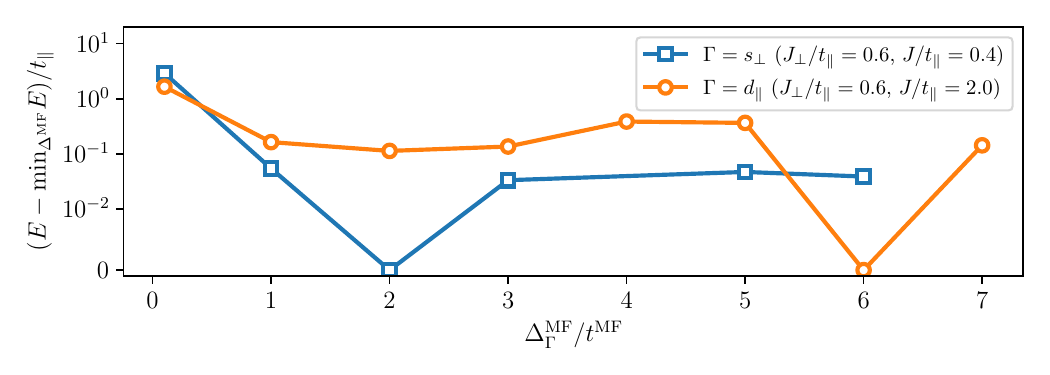}
\caption{Final ground state energies for $8\times 8\times 2$ mixD bilayers obtained with the G-HFPS for $\delta=0.5$, and different pairing symmetries as a function of the different pairing fields during the mean field (MF) optimization, $\Delta_{\Gamma}^{\mathrm{MF}}/t^{\mathrm{MF}}$.}
\label{fig:pairingfields}
\end{figure}

\begin{figure}[t]
\centering
\includegraphics[width=0.99\textwidth]{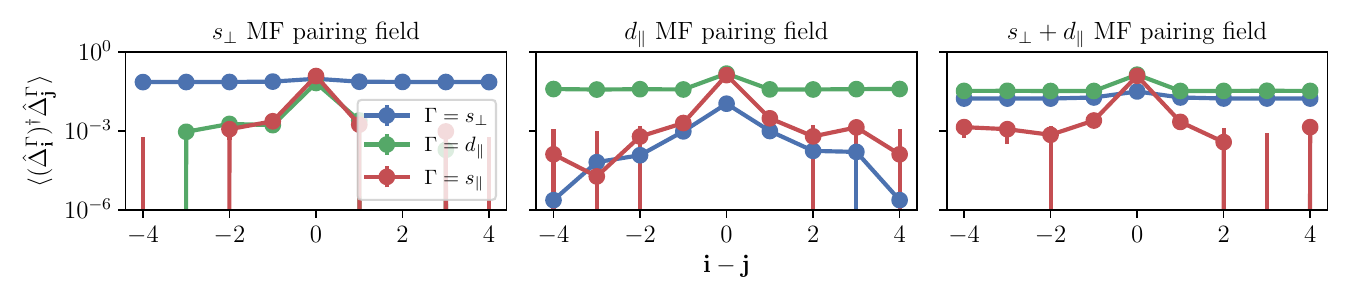}
\caption{Pairing correlations $\langle (\hat{\Delta}_\mathbf{i}^\Gamma)^\dagger\hat{\Delta}_\mathbf{j}^\Gamma\rangle$ with $\Gamma=s_\perp, d_\parallel, s_\perp+d_\parallel$ (blue, green,red markers) directly after the MF optimization with pairing fields $\Delta_{\Gamma^\prime}^{\mathrm{MF}}/t^{\mathrm{MF}}=2.0$ and $\Gamma^\prime=s_\perp, d_\parallel, s_\perp+d_\parallel$ (left to right).}
\label{fig:pairingfields_pcorrs}
\end{figure}

The mean-field solution is obtained by using exact gradient descent. In practice, we use $U^{\mathrm{MF}}/t^{\mathrm{MF}}=3.0$ and use constant pairing fields $\Delta_{\perp}^{\mathrm{MF}}$ ($\Delta_{\parallel}^{\mathrm{MF}}$) between (within) the layers or both. $\Delta_{_\perp}^{\mathrm{MF}}$ and $\Delta_{\parallel}^{\mathrm{MF}}$ are chosen according to the lowest energy obtained after the optimization of the full G-HFPS, initializing with the respective MF solution. Exemplary results for a $8\times 8\times 2$ bilayer at quarter-filling and different pairing fields $\Delta_{\perp}^{\mathrm{MF}}/t^{\mathrm{MF}}$  and $\Delta_{\parallel}^{\mathrm{MF}}/t^{\mathrm{MF}}$ are shown in Fig. \ref{fig:pairingfields} for two different regions in the phase diagram with interlayer $s$-wave and intralayer $d$-wave pairing, respectively. We always apply the respective $\Delta_{\perp(\parallel)}^{\mathrm{MF}}$, while setting the other $\Delta_{\parallel(\perp)}^{\mathrm{MF}}=0$. It can be seen that in both cases the energy obtained with a small pairing field $\Delta_{\perp, \parallel}^{\mathrm{MF}}/t^{\mathrm{MF}}=0.1$ yields a significantly higher energy than $\Delta_{\perp, \parallel}^{\mathrm{MF}}/t^{\mathrm{MF}}\geq 1.0$. 
The pairing correlations right after the MF optimization are shown in Fig.~\ref{fig:pairingfields_pcorrs}.

\FloatBarrier
\subsection{Number of parameters}
\begin{figure}[htp]
\centering
\includegraphics[width=0.85\textwidth]{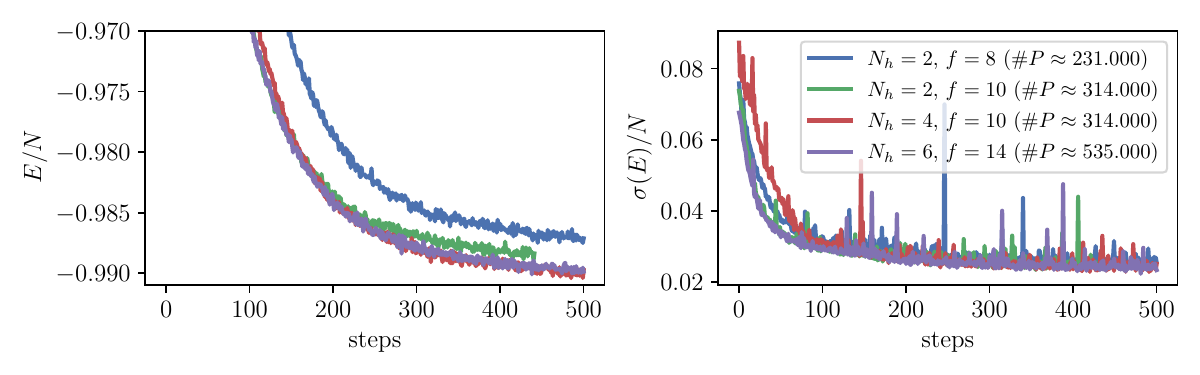}
\caption{Optimization curves for energy (left) and error (right) per site with different numbers of parameters $\#P$ from different numbers of hidden fermions $N_h$ and GCNN features $f$. We consider $J_\perp/t_\parallel=0.6$, $J_\parallel/t_\parallel=0.4$ and $\delta=0.5$.}
\label{fig:nparams}
\end{figure}
Throughout most of the work, we consider $N_h=4$ hidden fermions and a GCNN feature dimension of $f=10$. Fig.~\ref{fig:nparams} shows that increasing the number of parameters further does not improve the energy significantly for the considered 500 training steps.

\FloatBarrier
\subsection{Spatial Symmetries}
\begin{figure}[htp]
\centering
\includegraphics[width=0.85\textwidth]{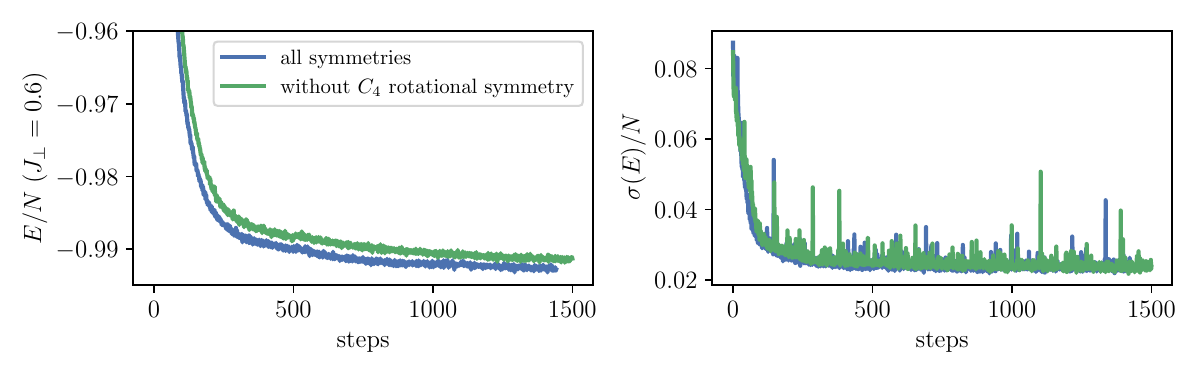}
\caption{Optimization curves for energy (left) and error (right) per site with all symmetries imposed (blue) and without the $C_4$ rotational symmetry (green). We consider $J_\perp/t_\parallel=0.6$, $J_\parallel/t_\parallel=0.4$ and $\delta=0.5$. To keep the number of parameters in both cases comparable, we consider the fully symmetrized G-HFPS with $N_h=4$ and $f=10$, and the G-HFPS without $C_4$ symmetry with $N_h=8$ and $f=18$.}
\label{fig:syms}
\end{figure}

In Fig. \ref{fig:syms}, we show the optimization curves with all symmetries imposed (blue) and without the $C_4$ rotational symmetry (green). The final energy increases by approx. $0.1\%$ relative to the full symmetrization when the rotational symmetry is not taken into account (decreasing the computational cost by a factor of $4$).  

\FloatBarrier
\subsection{Number of optimization steps}
\begin{figure}[htp]
\centering
\includegraphics[width=0.95\textwidth]{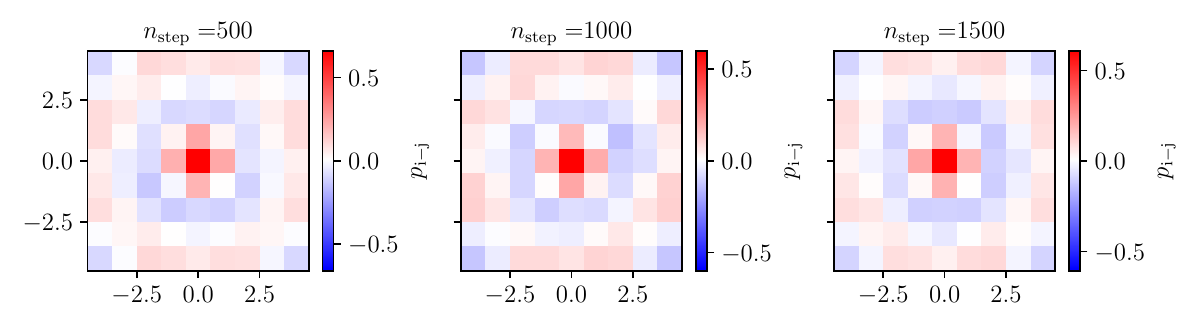}
\caption{The real-space mixed-estimator at $J_\perp/t_\parallel=0.6$, $J_\parallel/t_\parallel=0.4$ and $\delta=0.5$ after 500, 1000 and 1500 optimization steps (left to right).}
\label{fig:steps}
\end{figure}

If not stated differently, the G-HFPS considered in this work are trained for 1500 optimization steps. In the last 500 steps, the energy does not improve significantly. Furthermore, other observables like the real-space mixed-estimator do not change significantly after 500 optimization steps, see Fig. \ref{fig:steps}.

\FloatBarrier
\subsection{Convergence for different regions of the phase diagram}
\begin{figure}[t]
\centering
\includegraphics[width=0.95\textwidth]{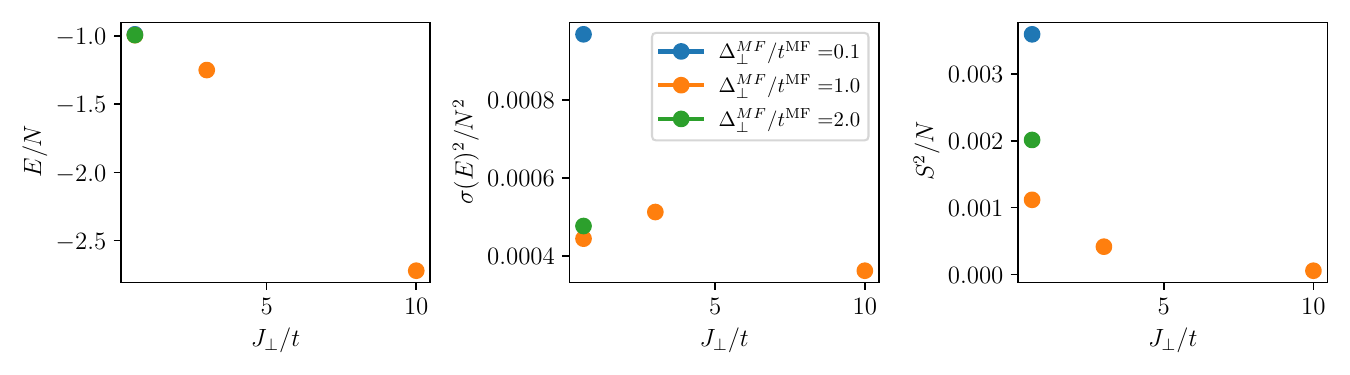}
\caption{Final ground state energies  (left), variance (middle) and total $S^2$ (right) per site $N$ for $4\times 4\times 2$ mixD bilayers obtained with the G-HFPS for $\delta=0.5$ and different parameters $J_\perp/t_\parallel =0.6, J_\parallel/t_\parallel =0.4$, $J_\perp/t_\parallel =3.0, J_\parallel/t_\parallel =0.0$ and $J_\perp/t_\parallel =10.0, J_\parallel/t_\parallel =0.0$, using different pairing fields $\Delta_{\perp}^{\mathrm{MF}}/t^{\mathrm{MF}}$ ($\Delta_{\parallel}^{\mathrm{MF}}=0.0$) for the first parameter set.}
\label{fig:phasediag_convergence}
\end{figure}

The convergence differs for different regions in the phase diagram, as we show exemplarily in Fig. \ref{fig:phasediag_convergence} for $4\times 4\times 2$ mixD bilayers with doping $\delta=0.5$: Comparing different parameters $J_\perp/t_\parallel =0.6, J_\parallel/t_\parallel =0.4$, $J_\perp/t_\parallel =3.0, J_\parallel/t_\parallel =0.0$ and $J_\perp/t_\parallel =10.0, J_\parallel/t_\parallel =0.0$, we find that convergence is enhanced for $J_\perp \gg J_\parallel$. This corresponds to the regime where the system can be mapped to a XXZ model (Heisenberg model for $\delta=0.5$).

\FloatBarrier
\subsection{System Size}
\begin{figure}[htp]
\centering
\includegraphics[width=0.95\textwidth]{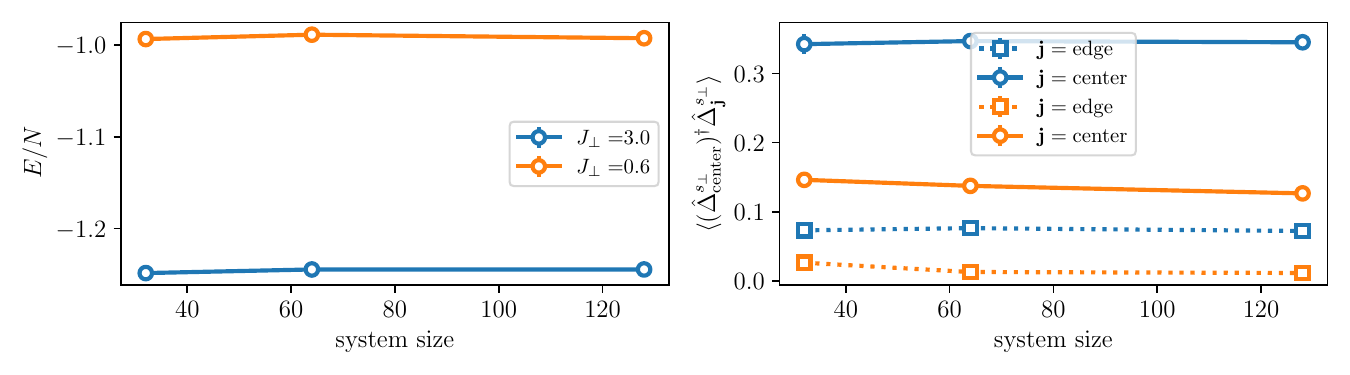}
\caption{Final ground state energies per site (left) and pairing correlations (right) for $J_\perp/t_\parallel =0.6, J_\parallel/t_\parallel =0.4$ (orange) and $J_\perp/t_\parallel =3.0, J_\parallel/t_\parallel =0.0$ (blue) and different system sizes corresponding to $N=4\times 4\times 2, 4\times 8\times 2, 8\times 8\times 2$ bilayers.}
\label{fig:sizes}
\end{figure}

In addition to the energies for different system sizes shown in Fig. \ref{fig:Fig1} in the main text (and reprinted in Fig. \ref{fig:sizes} (left)), further quantities can be compared as a function of the system sizes. Here, we additionally show the interlayer $s$-wave pairing correlations (see Fig. \ref{fig:sizes} (left and right) for two points in the mixD parameter space. All quantities are relatively constant as w.r.t. the system size. In Sec. \ref{sec:MixedEstimator}, also the system size scaling of the mixed estimator will be considered.

\FloatBarrier
\subsection{Comparison to MPS}
\begin{figure}[htp]
\centering
\includegraphics[width=0.95\textwidth]{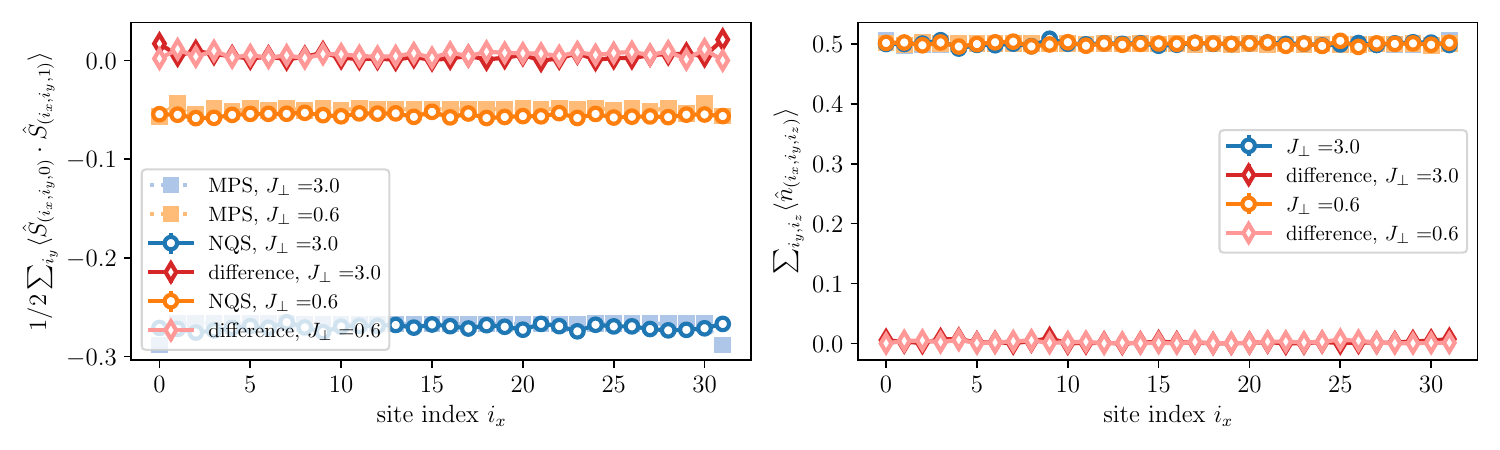}
\caption{Comparison of bond-spin correlations $1/2 \sum_{i_y} \langle \hat{S}_{(i_x,i_y,0)}\cdot \hat{S}_{(i_x,i_y,1)} \rangle$ (left) and densities $\sum_{i_y, i_z} \langle \hat{n}_{(i_x,i_y,i_z)}\rangle$ (right) from MPS and NQS and for the $32\times 2\times 2$ system at quarter-filling and with different parameters $J_\perp/t_\parallel, \, J_\parallel/t_\parallel$. Absolute differences are shown in red.}
\label{fig:MPScomparison}
\end{figure}
In addition to the comparison of energies from the G-HFPS (NQS) and MPS in \ref{fig:Fig1}b, we also compare the bond-spin correlations $1/2 \sum_{i_y} \langle \hat{S}_{(i_x,i_y,0)}\cdot \hat{S}_{(i_x,i_y,1)} \rangle$ and densities $\sum_{i_y, i_z} \langle \hat{n}_{(i_x,i_y,i_z)}\rangle$ in Fig. \ref{fig:MPScomparison} for the $32\times 2\times 2$ system. As in Fig. \ref{fig:Fig1}, we consider quarter-filling and different parameters $J_\perp/t_\parallel, \, J_\parallel/t_\parallel$. In all considered cases, we find very good agreement between the different methods, with the largest discrepancies at the boundaries due to the different boundary conditions that are applied in $x$-direction (open boundaries for MPS; periodic boundaries for NQS).

\FloatBarrier
\subsubsection{Convergence of MPS}
\begin{figure}[htp]
\centering
\includegraphics[width=0.85\textwidth]{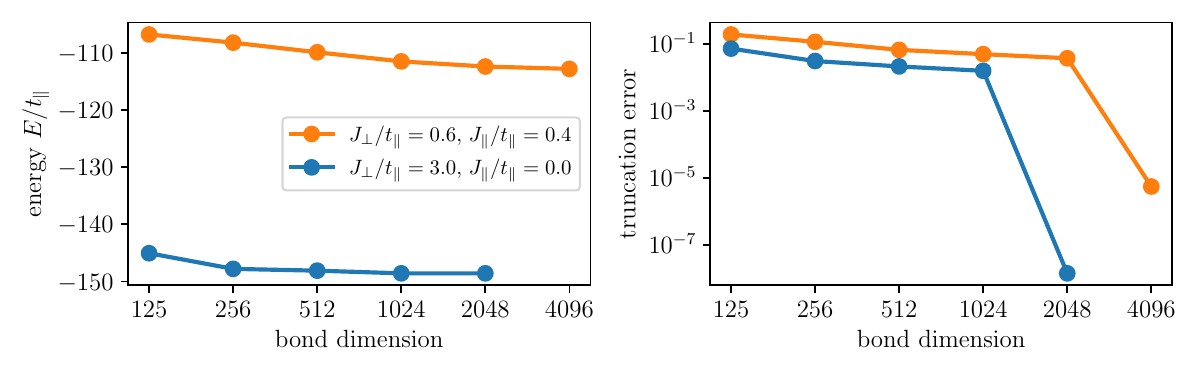}
\caption{Convergence of MPS calculations for two sets of Hamiltonian parameters. Left: Decreasing energy with the bond dimension. Right: Decreasing truncation error with the bond dimension.}
\label{fig:MPSconvergence}
\end{figure}

In Fig. \ref{fig:MPSconvergence}, we show the energy and truncation error of the MPS with increasing bond dimension. The final truncation error for $J_\perp/t_\parallel=3.0$ and $J_\parallel/t_\parallel=0.0$ (blue) is on the order of $10^{-8}$, for $J_\perp/t_\parallel=0.6$ and $J_\parallel/t_\parallel=0.4$ (orange) on the order of $10^{-6}$.

\FloatBarrier
\section{Density-density correlations} 
\begin{figure}[htp]
\centering
\includegraphics[width=1\textwidth]{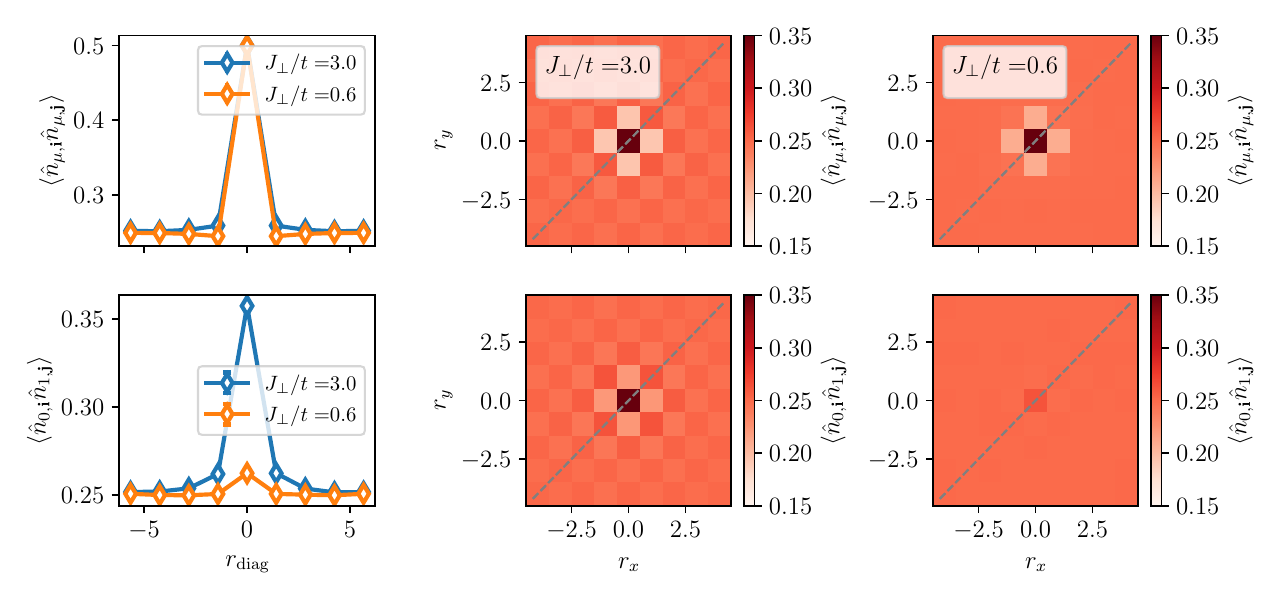}
\caption{Density-density correlations within the layers ($\langle \hat{n}_{\mu\mathbf{i}}\hat{n}_{\mu\mathbf{j}}\rangle$, top row) and between the layers ($\langle \hat{n}_{0\mathbf{i}}\hat{n}_{1\mathbf{j}}\rangle$, bottom row). For two exemplary points in the phase diagram, $J_\perp/t_\parallel =3.0, J_\parallel/t_\parallel =0.0$ and $J_\perp/t_\parallel =0.6, J_\parallel/t_\parallel =0.4$ at $\delta=0.5$, we show diagonal cuts on the leftmost panels in blue and orange, respectively. The corresponding density correlation maps are shown in the middle panel for $J_\perp/t_\parallel =3.0, J_\parallel/t_\parallel =0.0$ and on the right panel for $J_\perp/t_\parallel =0.6, J_\parallel/t_\parallel =0.4$.  }
\label{fig:densitycorrs}
\end{figure}

In Fig.~\ref{fig:densitycorrs}, we present density-density correlations within the layers ($\langle \hat{n}_{\mu\mathbf{i}}\hat{n}_{\mu\mathbf{j}}\rangle$, top row) and between the layers ($\langle \hat{n}_{0\mathbf{i}}\hat{n}_{1\mathbf{j}}\rangle$, bottom row). We consider two exemplary points in the phase diagram, $J_\perp/t_\parallel =3.0, J_\parallel/t_\parallel =0.0$ and $J_\perp/t_\parallel =0.6, J_\parallel/t_\parallel =0.4$, both at $\delta=0.5$, after $500$ optimization steps and exploiting the the full $C_4$ symmetry of the system. The diagonal cuts reveal a stronger signal for interlayer density correlations when $J_\perp$ is increased. The full density correlation maps for the corresponding $J_\perp/t_\parallel =3.0, J_\parallel/t_\parallel =0.0$ in the middle panel reveal relatively long-range signals in both inter- and intralayer density correlations, with a minimal value for nearest-neighbors. Decreasing $J_\perp$ (right panel) this long-range signal vanishes, especially for the interlayer correlations.

\section{Mixed momentum-real space estimator \label{sec:MixedEstimator}}
\subsection{Approximating the pairing order parameter}
In the main text, we consider the following pairing symmetries
\begin{equation} 
\begin{aligned}
    \hat{\Delta}^{s_\perp}_\mathbf{l}&\propto \left(\hat{c}_{\mathbf{l},0,\downarrow}\hat{c}_{\mathbf{l},1,\uparrow} -\hat{c}_{\mathbf{l},0,\uparrow}\hat{c}_{\mathbf{l},1,\downarrow} \right)\\
\hat{\Delta}^{d_\parallel}_{\mu,\mathbf{l}}
&\propto\sum_{\delta=\pm\hat{x},\pm\hat{y}}{\eta_\delta^{d}}
\left(\hat{c}_{\mathbf{l},\mu,\downarrow}\hat{c}_{\mathbf{l}+\delta,\mu,\uparrow}
-\hat{c}_{\mathbf{l},\mu,\uparrow}\hat{c}_{\mathbf{l}+\delta,\mu,\downarrow}\right).
\label{eq:Deltas_SM}
\end{aligned}
\end{equation}

In order to study the pairing properties of the system, we would like to calculate the order parameter 
\begin{align}
    \Delta^{s_\perp}_{\mathbf{k}} = \langle \hat{c}_{\mathbf{k},0,\uparrow}\hat{c}_{\mathbf{-k},1,\downarrow}\rangle\quad \mathrm{and} \quad \Delta^{d_\parallel}_{\mathbf{k},\mu} = \langle \hat{c}_{\mathbf{k},\mu,\uparrow}\hat{c}_{\mathbf{-k},\mu,\downarrow}\rangle
\end{align}
or the full pair-pair map
\begin{align}
    M_{\mathbf{k,k^\prime}}^{s_\perp} = \langle \hat{c}^\dagger_{\mathbf{-k},1,\downarrow}\hat{c}^\dagger_{\mathbf{k},0,\uparrow}\hat{c}_{\mathbf{k^\prime},0,\uparrow}\hat{c}_{\mathbf{-k^\prime},1,\downarrow}\rangle \quad \mathrm{and}\quad M_{\mathbf{k,k^\prime},\mu}^{d_\parallel} = \langle \hat{c}^\dagger_{\mathbf{-k},\mu,\downarrow}\hat{c}^\dagger_{\mathbf{k},\mu,\uparrow}\hat{c}_{\mathbf{k^\prime},\mu,\uparrow}\hat{c}_{\mathbf{-k^\prime},\mu,\downarrow}\rangle.
\end{align}
In the following discussion, we absorb the layer index $\mu$ into $\mathbf{k}$ for the intra-layer pairing.
For finite-size systems and fixed particle number, we face the problem that $\Delta^\Gamma_{\mathbf{k}}=0$ even in the presence of superconductivity. For $M_{\mathbf{k,k^\prime}}$ the problem is that it involves $(L_xL_y)^4$ terms and is too expensive to evaluate. In order to still gain insights into these quantities, we consider the mixed-estimator defined in Eq.~\eqref{eq:MixedEstimator} in the main text and reprinted here for clarity:
\begin{align}
\langle \hat{P}^{s_\perp}(\mathbf{k})\rangle  &=\frac{1}{ \langle \bar{\Delta}^\dagger \bar{\Delta}\rangle^{1/2}}\langle \hat{c}_{\mathbf{k},1,\uparrow}^\dagger\hat{c}_{-\mathbf{k},0,\downarrow}^\dagger \hat{\Delta}_{\mathbf{l}}^{s_\perp}\rangle= \frac{1}{ \langle \bar{\Delta}^\dagger \bar{\Delta}\rangle^{1/2}}\frac{1}{\sqrt{2}N}\sum_{\mathbf{i,j}}e^{i\mathbf{k}(\mathbf{i}-\mathbf{j})}\langle \hat{c}_{\mathbf{i},1,\uparrow}^\dagger\hat{c}_{\mathbf{j},0,\downarrow}^\dagger \hat{\Delta}^{s_\perp}_\mathbf{l}\rangle.\label{eq:mixedEstimator_sm} \\
\langle \hat{P}^{d_\parallel}(\mathbf{k})\rangle &=\frac{1}{ \langle \bar{\Delta}^\dagger \bar{\Delta}\rangle^{1/2}}\langle \hat{c}_{\mathbf{k},\uparrow}^\dagger\hat{c}_{-\mathbf{k},\downarrow}^\dagger \hat{\Delta}_{\mathbf{l}}^{d_\parallel}\rangle = \frac{1}{ \langle \bar{\Delta}^\dagger \bar{\Delta}\rangle^{1/2}}\frac{1}{N}\sum_{\mathbf{i,j}}e^{i\mathbf{k}(\mathbf{i}-\mathbf{j})}\langle \hat{c}_{\mathbf{i},\uparrow}^\dagger\hat{c}_{\mathbf{j},\downarrow}^\dagger \hat{\Delta}_{\mathbf{l}}^{d_\parallel}\rangle.  
\end{align}
In the following, we consider the specific case of interlayer $s$-wave pairing but the discussion can be straightforwardly adapted to the intralayer pairing. For simplicity, we absorb the layer indices $0,1$ into the site indices $\mathbf{i,j}$ and omit the symmetry label $\Gamma$.  \\

In what follows, we will motivate the use of \eqref{eq:mixedEstimator_sm}: In the presence of superconductivity, we expect that pair creation and annihilation processes become increasingly uncorrelated as the system size increases. This implies that $M$ is dominated by a single eigenvector
\begin{align}
    M_{\mathbf{k,k^\prime}} = P_{\mathbf{k}}P_{\mathbf{k^\prime}}+\mathcal{O}(1/N).
    \label{eq:Meigenvecs}
\end{align}

The eigenvector $P_{\mathbf{k}}$ can be extracted by the power method: In order to do so, we define we define $g_\mathbf{k}=\mathrm{cos}k_z$ (for the inter-layer $s$-wave pairing) and multiply Eq.~\eqref{eq:Meigenvecs} with the vector $\mathbf{g}=(g_\mathbf{k})_\mathbf{k}$:
\begin{align}
    \frac{1}{N}\sum_{\mathbf{k^\prime}}M_{\mathbf{k,k^\prime}}g_{\mathbf{k^\prime}}\approx \frac{1}{N}P_\mathbf{k}(\mathbf{g}\cdot \mathbf{P}).
\end{align}
To get $P_\mathbf{k}$ we note that
\begin{align}
    \left(\frac{\mathbf{g}\cdot \mathbf{P}}{N}\right)^2 = \frac{1}{2} \langle \bar{\Delta}^\dagger \bar{\Delta}\rangle. 
\end{align}
Both together gives
\begin{align}
    \frac{1}{N}\sum_{\mathbf{k^\prime}}M_{\mathbf{k,k^\prime}}g_{\mathbf{k^\prime}}\approx \frac{1}{N}P_\mathbf{k}\sqrt{\frac{1}{2} \langle \bar{\Delta}^\dagger \bar{\Delta}\rangle}
\end{align}
and hence
\begin{equation}
\begin{aligned}
     \langle \hat{P}(\mathbf{k})\rangle:= P(\mathbf{k} )&= \frac{\sqrt{2}}{N \langle \bar{\Delta}^\dagger \bar{\Delta}\rangle^{1/2}}\sum_{\mathbf{k^\prime}}M_{\mathbf{k,k^\prime}}g_{\mathbf{k^\prime}}\\
     &=\frac{\sqrt{2}}{N \langle \bar{\Delta}^\dagger \bar{\Delta}\rangle^{1/2}}\sum_{\mathbf{k^\prime}}g_{\mathbf{k^\prime}}\langle \hat{c}^\dagger_{\mathbf{-k}\downarrow}\hat{c}^\dagger_{\mathbf{k}\uparrow}\hat{c}_{\mathbf{k^\prime}\uparrow}\hat{c}_{\mathbf{-k^\prime}\downarrow}\rangle \\
    &=
    \frac{1}{ \langle \bar{\Delta}^\dagger \bar{\Delta}\rangle^{1/2}}\langle \hat{c}^\dagger_{\mathbf{-k}\downarrow}\hat{c}^\dagger_{\mathbf{k}\uparrow}\bar{\Delta}\rangle \\ 
    &=\frac{1}{ \langle \bar{\Delta}^\dagger \bar{\Delta}\rangle^{1/2}}\frac{1}{N}\sum_{\mathbf{i,j}}e^{i\mathbf{k}(\mathbf{i}-\mathbf{j})} \langle \hat{p}({\mathbf{i}-\mathbf{j}})\rangle
\end{aligned}
\end{equation}
where we have defined
\begin{align}
    \langle \hat{p}({\mathbf{i}-\mathbf{j}})\rangle = \langle\hat{c}^\dagger_{\mathbf{i}\downarrow}\hat{c}^\dagger_{\mathbf{j}\uparrow}\bar{\Delta}\rangle.
    \label{eq:mixedEstimator_real}
\end{align}
If we assume $\bar{\Delta}\approx\langle \hat{\Delta}_\mathbf{l}\rangle$ for every $\mathbf{l}$, this exactly corresponds to Eq. \eqref{eq:mixedEstimator_sm}.\\

Furthermore, we make use of
\begin{equation}
    \begin{aligned}
    \hat{\bar{\Delta}}=\frac{1}{N}\sum_{\mathbf{i}}\hat{\Delta}_\mathbf{i} &=\frac{1}{\sqrt{2}N}\sum_{\mathbf{i}}(\hat{c}_{\mathbf{i}\uparrow}\hat{c}_{\mathbf{i+e}_z\downarrow}-\hat{c}_{\mathbf{i}\downarrow}\hat{c}_{\mathbf{i+e}_z\uparrow}) \\
    &=\frac{1}{\sqrt{2}N^2}\sum_{\mathbf{k}\mathbf{k^\prime}}\sum_{\mathbf{i},\mathbf{j}}e^{i(\mathbf{k}+\mathbf{k^\prime})\cdot  \mathbf{i}+i\mathbf{k^\prime}\cdot \mathbf{e}_z}(\hat{c}_{\mathbf{k}\uparrow}\hat{c}_{\mathbf{k^\prime}\downarrow}-\hat{c}_{\mathbf{k}\downarrow}\hat{c}_{\mathbf{k^\prime}\uparrow})
    \\
    &=\frac{1}{\sqrt{2}N}\sum_\mathbf{k}e^{-i\mathbf{k}\cdot \mathbf{e}_z}\left(\hat{c}_{\mathbf{k}\uparrow}\hat{c}_{\mathbf{-k}\downarrow}-\hat{c}_{\mathbf{k}\downarrow}\hat{c}_{\mathbf{-k}\uparrow}\right)
     \\
    &=\frac{\sqrt{2}}{N}\sum_\mathbf{k}\mathrm{cos}k_z\hat{c}_{\mathbf{k}\uparrow}\hat{c}_{\mathbf{-k}\downarrow} \\
    &=\frac{\sqrt{2}}{N}\sum_\mathbf{k}g_\mathbf{k}\hat{c}_{\mathbf{k}\uparrow}\hat{c}_{\mathbf{-k}\downarrow}.
    \end{aligned}
\end{equation}
In practice, we also subtract the mean-field part:
        \begin{align}
            \langle\!\langle \hat{c}_{\mathbf{k,}0,\uparrow}^\dagger\hat{c}_{\mathbf{k},1,\downarrow}^\dagger \hat{\Delta}_{\mathbf{l}}\rangle\!\rangle = \langle \hat{c}_{\mathbf{k,}0,\uparrow}^\dagger\hat{c}_{\mathbf{k},1,\downarrow}^\dagger \hat{\Delta}_{\mathbf{l}}\rangle-\mathrm{Wick}\,\mathrm{contractions},
        \end{align}
where only $\langle \hat{c}_{\mathbf{i},0,\uparrow}^\dagger\hat{c}_{\mathbf{l},0,\uparrow} \rangle \langle \hat{c}_{\mathbf{j},1,\downarrow}^\dagger \hat{c}_{\mathbf{l},1,\downarrow}\rangle$ contributes with a significant magnitude. 

\subsection{Finite size scaling}
\begin{figure}[t]
\centering
\includegraphics[width=0.75\textwidth]{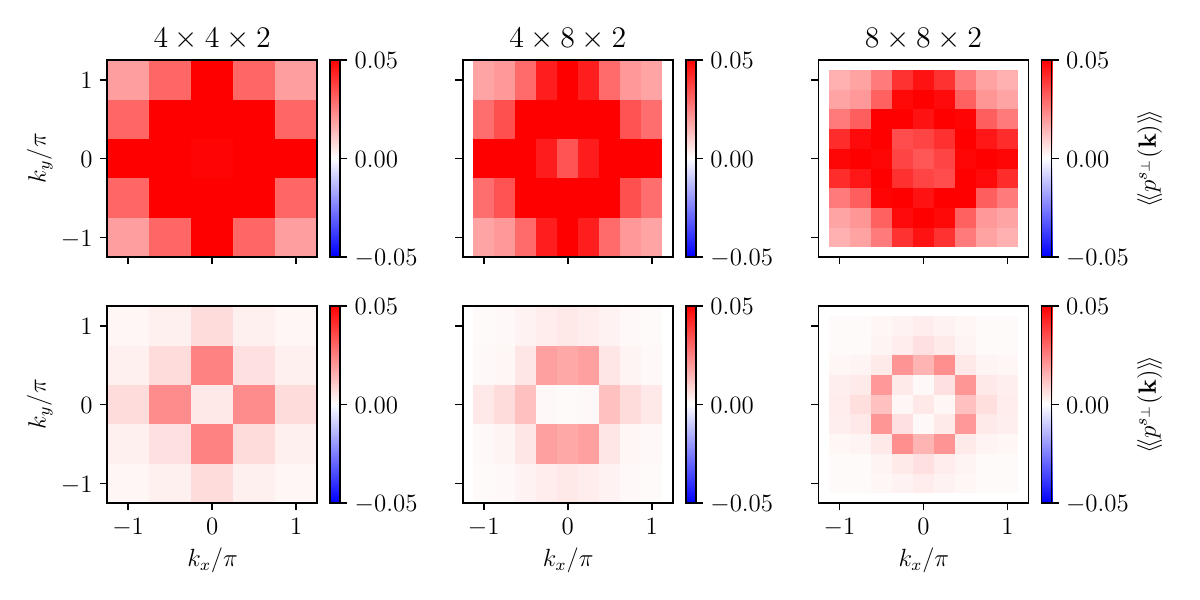}
\caption{Mixed estimators for $J_\perp/t_\parallel =3.0, J_\parallel/t_\parallel =0.0$ (top row) and $J_\perp/t_\parallel =0.6, J_\parallel/t_\parallel =0.4$ (bottom row) and different system sizes corresponding to $N=4\times 4\times 2, 4\times 8\times 2, 8\times 8\times 2$ bilayers (left to right).}
\label{fig:MEsizes}
\end{figure}

In Fig. \ref{fig:MEsizes}, the mixed estimator is shown for different system sizes $N=4\times 4\times 2, 4\times 8\times 2, 8\times 8\times 2$ and $\delta=0.5$.

\FloatBarrier
\subsection{Estimating the size of the pairs: mixed-estimator in real space}
\begin{figure}[htp]
\centering
\includegraphics[width=0.99\textwidth]{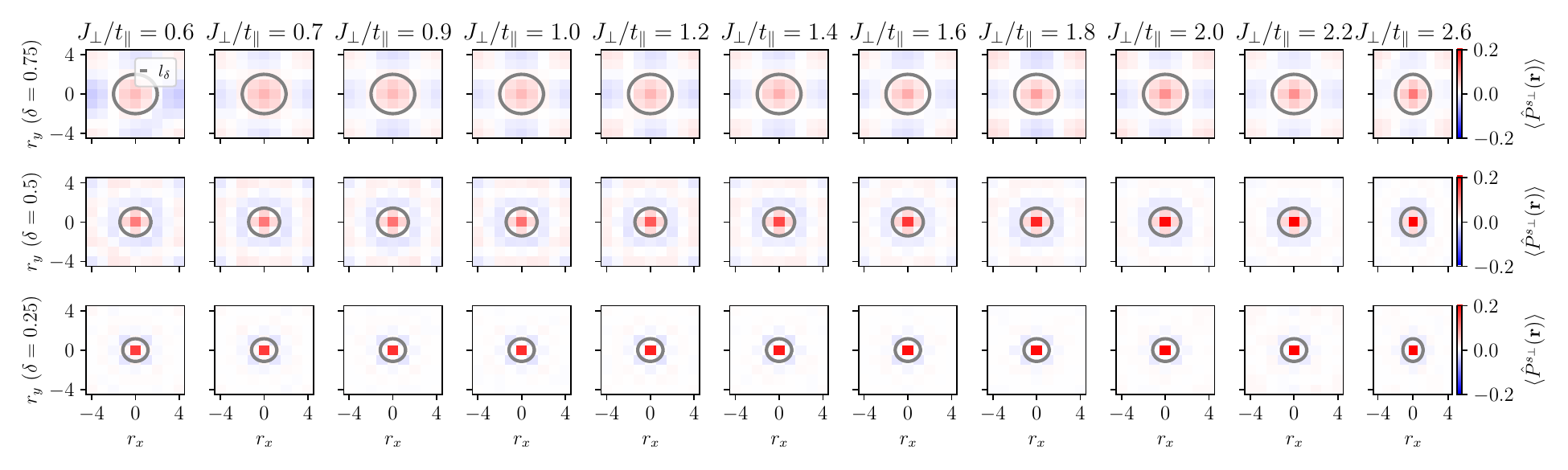}
\caption{interlayer $s_\perp$-wave mixed estimator in real space for different $J_\perp/t_\parallel$ and hole doping $\delta=0.75$ (top), $\delta=0.5$ (middle) and $\delta=0.25$ (bottom). }
\label{fig:MEreal}
\end{figure}

\begin{figure}[htp]
\centering
\includegraphics[width=0.99\textwidth]{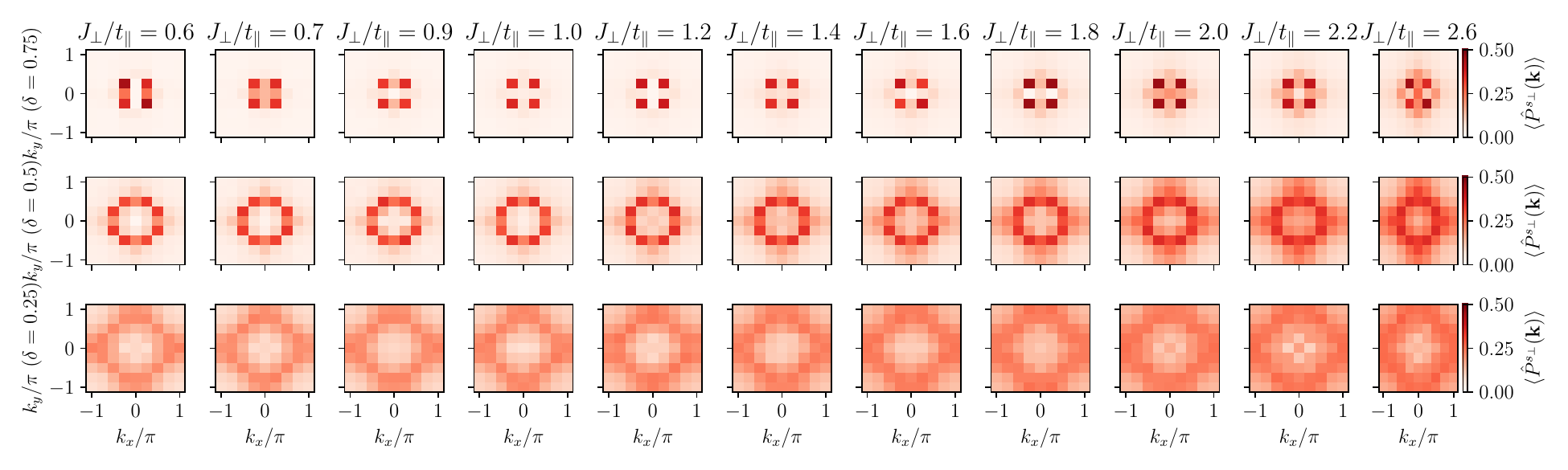}
\caption{interlayer $s_\perp$-wave mixed estimator in momentum space for different $J_\perp/t_\parallel$ and hole doping $\delta=0.75$ (top), $\delta=0.5$ (middle) and $\delta=0.25$ (bottom). }
\label{fig:MEmom}
\end{figure}

To get insights into the size of the pairs when tuning over the BEC to BCS transition, we transform the mixed-estimator back to real space, see Eq.~\eqref{eq:mixedEstimator_real}, resulting in Fig.~\ref{fig:MEreal}. For comparison, we also show the mixed-estimator in momentum space, see Fig.~\ref{fig:MEmom}.

\FloatBarrier
\FloatBarrier
\section{Additional results on the BEC-to-BCS crossover}
\begin{figure}[htp]
\centering
\includegraphics[width=0.5\textwidth]{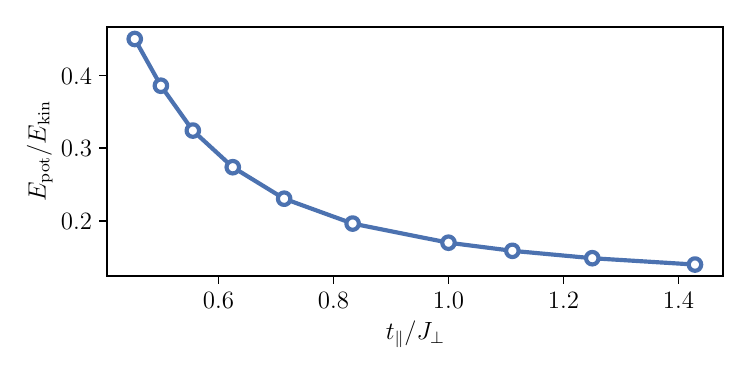}
\caption{The fraction of potential and kinetic energies, $E_\mathrm{pot}/E_\mathrm{kin}$ as a function of $t_\parallel/J_\perp$ at $\delta=0.5$ and $J_\parallel/t_\parallel=0.4$.
}
\label{fig:EpotvsEkin}
\end{figure}

In addition to the results on the BEC-to-BCS crossover in the main text, we show in Fig.~\ref{fig:EpotvsEkin} the fraction of potential and kinetic energies, $E_\mathrm{pot}/E_\mathrm{kin}$ as a function of $t_\parallel/J_\perp$ at $\delta=0.5$ and $J_\parallel/t_\parallel=0.4$, where
\begin{align}
    E_\mathrm{kin} = -\langle \sum_{\langle \mathbf{i,j} \rangle, \mu, \sigma} 
 \mathcal{\hat{P}}_G\left( \hat{c}^{\dagger}_{\mathbf{i},\mu,\sigma} \hat{c}_{\mathbf{j},\mu,\sigma} + \text{h.c.} \right)\rangle
\end{align}
and $E_\mathrm{pot}=E-E_\mathrm{kin}$. The ratio $E_\mathrm{pot}/E_\mathrm{kin}\lesssim 0.5$ throughout the considered regime of $t_\parallel/J_\perp$, i.e. the system is dominated by the kinetic energy. This is probably due to the relatively large doping $\delta=0.5$. Furthermore, $E_\mathrm{pot}/E_\mathrm{kin}$ takes its largest value in the BEC regime and decreases to a value of $E_\mathrm{pot}/E_\mathrm{kin}\approx 0.1$ when crossing over to the BCS regime. This behavior is expected for a BEC-to-BCS crossover~\cite{Quijin2024superconductivitycrossesover}.
\section{Additional results for the different pairing symmetries}

\begin{figure}[htp]
\centering
\includegraphics[width=0.5\textwidth]{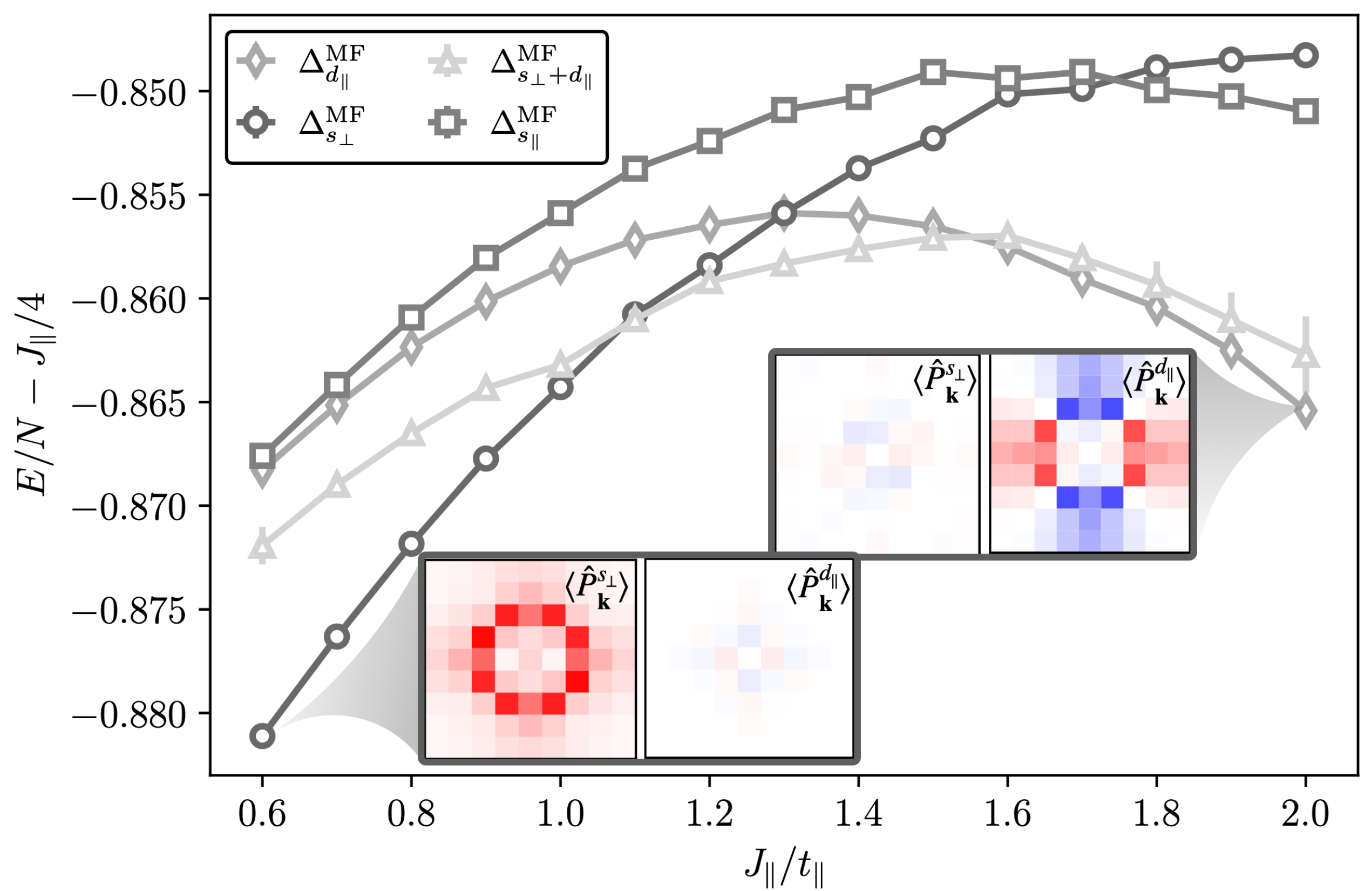}
\caption{Ground state energies obtained with different initializations using different pairing field symmetries and strengths. Initializing with the $s_\perp, s_\perp+d_\parallel, d_\parallel$ mean-field pairing fields leads to the lowest energies in the considered regime of $J_\parallel/t_\parallel$ and fixed $J_\perp/t_\parallel=0.6$.}
\label{fig:dsenergies}
\end{figure}

\begin{figure}[t]
\centering
\includegraphics[width=0.9\textwidth]{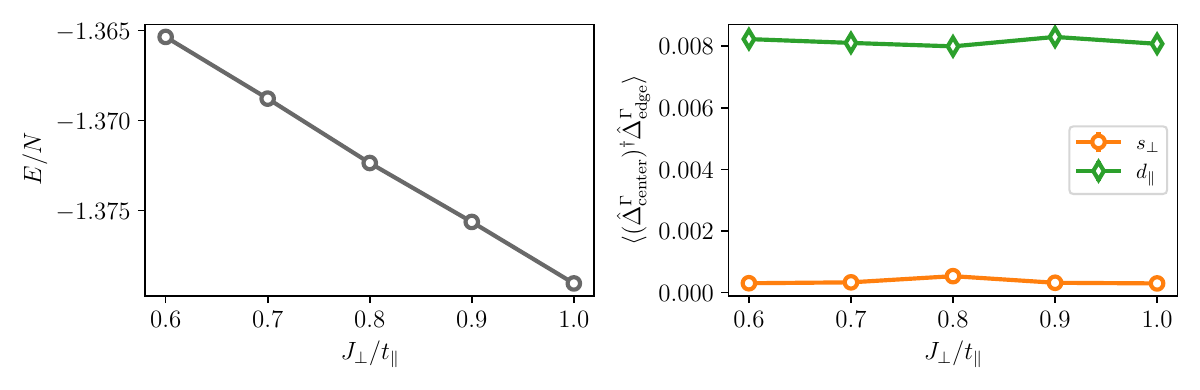}
\caption{Energy per site (left) and long-range $s_\perp$ (orange) and $d_\parallel$ (green) pairing correlations (right) when changing $J_\perp$ for fixed $J_\parallel/t_\parallel=2.0$.}
\label{fig:dsenergies2}
\end{figure}

In Fig.~\ref{fig:dsenergies}, we show the energies obtained with initializations that use different MF pairing fields, namely Eq.~\eqref{eq:Deltas_SM}. Initializing with the $s_\perp, s_\perp+d_\parallel, d_\parallel$ mean-field pairing fields leads to the lowest energies in the considered regime of $J_\parallel/t_\parallel$ and fixed $J_\perp/t_\parallel=0.6$. The $s_\parallel$ initialization leads higher energies throughout the considered regime of $J_\parallel/t_\parallel$.\\

Furthermore, we consider the enery and long-range pair correlations at fixed $J_\parallel/t_\parallel=2.0$ when changing $J_\perp/t_\parallel$ in Fig.~\ref{fig:dsenergies2}. For the considered regime of $J_\perp/t_\parallel=0.6,\dots,1.0$ we do not observe any change in the pairing behavior.

\section{The strong $J_\perp$ limit of the mixed-dimensional model \label{appendix:XXZ}}

\subsection{The mapping}
As shown in Ref.~\cite{Bohrdt2021}, the mixed-dimensional $t$-$J$ bilayer can be mapped to an effective spin-$1/2$ description when $J_{\perp} \gg t_{\parallel}, J_{\parallel}$.
In this limit, the low-energy sector of the Hamiltonian~\eqref{eq:mixDtJ} is restricted to two rung states at each lattice site $\mathbf{i}$: The first corresponds to a pair of holes residing on the two layers,
$\ket{0}_{\mathbf{i}} = \ket{0}_{\mathbf{i},0}\ket{0}_{\mathbf{i},1}$, while the second is an interlayer spin singlet,
$\ket{1}_{\mathbf{i}} = \hat{b}^{\dagger}_{\mathbf{i}}\ket{0}_{\mathbf{i}}$ created by the operator
\begin{align}
    \hat{b}^{\dagger}_{\mathbf{i}} \ket{0}_{\mathbf{i}} = \frac{1}{\sqrt{2}}\left( \hat{c}^{\dagger}_{\mathbf{i},\uparrow, 0} \hat{c}^{\dagger}_{\mathbf{i},\downarrow,1} - \hat{c}^{\dagger}_{\mathbf{i},\downarrow,0} \hat{c}^{\dagger}_{\mathbf{i},\uparrow,1} \right) \ket{0}_{\textbf{i}},
\end{align}
and subject to the hard-core constraint $\hat b_{\mathbf{i}}^{\dagger}\hat b_{\mathbf{i}}\le 1$.

Carrying out a second-order perturbative expansion in the intralayer hopping, following Ref.~\cite{Bohrdt2021} and related works~\cite{lange2024feshbachladder,lange2024pairingdome,schloemer2023superconductivity}, one finds that virtual hopping processes generate an effective hard-core boson model on a single square lattice.
The emergent parameters are an effective boson hopping
$t_\mathrm{eff} = 2t_{\parallel}^2/J_{\perp}$,
a nearest-neighbor density interaction
$V_\mathrm{eff} = 4t_{\parallel}^2/J_{\perp} - J_{\parallel}/2$,
and a chemical potential term
$\mu_b = -J_{\perp} - 2t_{\parallel}^2 z/J_{\perp}$,
where $z$ denotes the coordination number of the lattice ($z=2$ for the bilayer geometry).
Neglecting constant energy shifts, the effective Hamiltonian takes the form
\begin{equation}
\begin{aligned}
    \hat{\mathcal{H}}_{\mathrm{eff}} = &-t_\mathrm{eff} \sum_{\langle\mathbf{i}, \mathbf{j}\rangle} \hat{\mathcal{P}} \big(\hat{b}_{\mathbf{i}}^{\dagger} \hat{b}_{\mathbf{j}}^{\vphantom\dagger} + \text{h.c.} \big)\hat{\mathcal{P}} +\mu_b \sum_{\mathbf{i}} \hat{b}_{\mathbf{i}}^{\dagger} \hat{b}_{\mathbf{i}}^{\vphantom\dagger}  + V_\mathrm{eff} \sum_{\langle\mathbf{i}, \mathbf{j}\rangle} \hat{b}_{\mathbf{i}}^{\dagger} \hat{b}_{\mathbf{i}}^{\vphantom\dagger} \hat{b}_{\mathbf{j}}^{\dagger} \hat{b}_{\mathbf{j}}^{\vphantom\dagger}\, .
\end{aligned}
\label{eq:Hhcb}
\end{equation}
An effective spin representation is obtained by identifying the two local bosonic configurations with the down and up states of a spin-$1/2$ system.
Defining spin operators $\hat J_{\mathbf{i}}^{\mu}$ ($\mu=x,y,z$) via
\begin{equation}
    \begin{gathered}
        \hat{J}_{\mathbf{i}}^{+} = (-1)^{\mathbf{i}} \hat{b}^{\dagger}_{\mathbf{i}}, \quad  
         \hat{J}_{\mathbf{i}}^{-} = (-1)^{\mathbf{i}} \hat{b}_{\mathbf{i}}, \quad
         \hat{J}^z_{\mathbf{i}} = \hat{b}^{\dagger}_{\mathbf{i}} \hat{b}_{\mathbf{i}} - 1/2,
    \end{gathered}
    \label{eq:Jops}
\end{equation}
with $(-1)^{\mathbf{i}}=+1$ ($-1$) on the A (B) sublattice, the hard-core boson Hamiltonian~\eqref{eq:Hhcb} is mapped onto an XXZ spin model,
\begin{equation}
\begin{aligned}
    \hat{\mathcal{H}}_{\mathrm{XXZ}} = K \sum_{\langle\mathbf{i}, \mathbf{j}\rangle}\left( \hat{J}^x_{\mathbf{i}} \hat{J}^x_{\mathbf{j}} + \hat{J}^y_{\mathbf{i}} \hat{J}^y_{\mathbf{j}} + \delta \hat{J}^z_{\mathbf{i}} \hat{J}^z_{\mathbf{j}} \right)+\mathrm{const.}\,,
\end{aligned}
\label{eq:XXZ1}
\end{equation}
where $K = 4t_{\parallel}^2/J_{\perp}$ and $\delta = 1 - J_{\parallel}/(2K)$. The magnetization $m$ of the XXZ model~\eqref{eq:XXZ1}, is related to the particle density of the original bilayer by $m=n-\tfrac12$.
For vanishing intralayer exchange $J_{\parallel}=0$ and quarter filling, the model reduces to the 2D Heisenberg Hamiltonian.

\subsection{Numerical signatures of the XXZ limit}
\begin{figure}[t]
\centering
\includegraphics[width=0.65\textwidth]{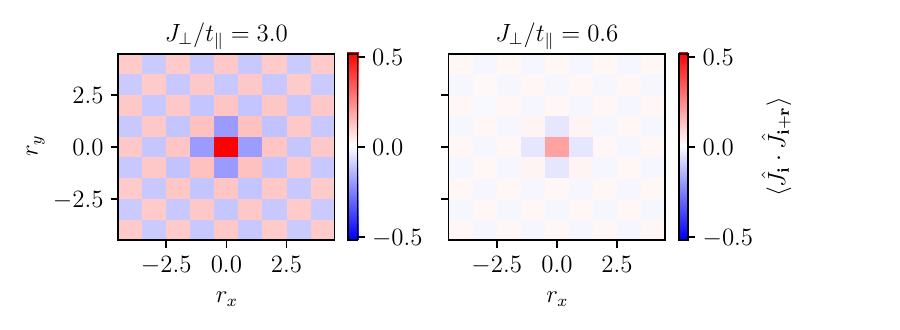}
\caption{The effective Heisenberg spin correlations $\langle \hat{J}_{\mathbf{i}}\cdot \hat{J}_{\mathbf{i}+\mathbf{r}}\rangle$ for the mixD model at quarter filling and different interlayer couplings $J_\perp$.}
\label{fig:XXZ}
\end{figure}

We note that using Eq.~\eqref{eq:Jops},
\begin{align}
    (\hat{\Delta}_\mathbf{i}^{s_\perp})^\dagger\hat{\Delta}_\mathbf{j}^{s_\perp}=\hat{b}_\mathbf{i}^\dagger \hat{b}_\mathbf{j} = (-1)^\mathbf{i}(-1)^\mathbf{j}\hat{J}_{\mathbf{i}}^{+}\hat{J}_{\mathbf{j}}^{-}.
\end{align}
For the Heiseinberg limit with $m=0$ and emerging $\mathrm{SU(2)}$ symmetry (i.e. $\langle \hat{J}_{\mathbf{i}}^{+}\hat{J}_{\mathbf{j}}^{-}\rangle =2 \langle\hat{J}^x_{\mathbf{i}} \hat{J}^{x}_{\mathbf{j}}\rangle= 2 \langle\hat{J}^y_{\mathbf{i}} \hat{J}^{y}_{\mathbf{j}}\rangle= 2 \langle\hat{J}^z_{\mathbf{i}} \hat{J}^{z}_{\mathbf{j}}\rangle$), this can be used to connect to the spin-spin correlator to the pair-pair correlations in the mixD bilayer,
\begin{align}
    \hat{J}_{\mathbf{i}}\cdot \hat{J}_{\mathbf{j}}=\frac{3}{2}(-1)^\mathbf{i}(-1)^\mathbf{j}(\hat{\Delta}_\mathbf{i}^{s_\perp})^\dagger\hat{\Delta}_\mathbf{j}^{s_\perp}.
\end{align}
In Fig.~\ref{fig:XXZ}, it can be seen that for strong $J_\perp$ the system features effective spin correlations $\langle \hat{J}_{\mathbf{i}}\cdot \hat{J}_{\mathbf{i}+\mathbf{r}}\rangle$ with long-range AFM order, as expected for the Heisenberg model.

\end{document}